\documentclass[%
 reprint,
preprintnumbers,
 amsmath,amssymb,
 aps,nofootinbib,
]{revtex4-2}
\usepackage{float} %placing figures
\usepackage{xcolor}
\usepackage{graphicx}% Include figure files
\usepackage{dcolumn}% Align table columns on decimal point
\usepackage{bm}% bold math
\usepackage{hyperref}% add hypertext capabilities
\hypersetup{backref, colorlinks=true  } %remove boxed in hyperref

\newcommand{\ep}{\epsilon}

\newcommand{\va}{{\mathbf{a}}}

\newcommand{\vk}{{\mathbf{k}}}
\newcommand{\hvk}{{\mathbf{\hat{\mathbf{k}}}}}
\newcommand{\vp}{{\mathbf{p}}}

\newcommand{\hvz}{{\hat{\mathbf{z}}}}

\newcommand{\cE}{{\cal{E}}}
\newcommand{\bcE}{{\bar{\cal{E}}}}

\newcommand{\la}{\langle}
\newcommand{\ra}{\rangle}

\newcommand{\nn}{\nonumber}

\newcommand{\jo}[1]{\textcolor{black}{{#1}}} 
\DeclareMathOperator\arctanh{arctanh}

\begin{document}

\title{Ultracold spin-balanced fermionic quantum liquids with renormalized $P$-wave interactions}

\author{J.~M.~Alarc\'on}
\email{jmanuel.alarcon@uah.es}

\affiliation{ Universidad de Alcal\'a, Grupo de F\'{\i}sica Nuclear y de Part\'{\i}culas, Departamento de F\'{\i}sica y Matem\'aticas,  28805 Alcal\'a de Henares (Madrid), Spain}

\author{J.~A. Oller}
\email{oller@um.es}
\affiliation{Departamento de F\'{\i}sica, Universidad de Murcia, E-30071 Murcia, Spain.}

\date{\today}

\begin{abstract}
  We consider a spin-balanced degenerate gas of spin-1/2 fermions  whose dynamics is governed by low-energy $P$-wave interactions, characterized by the scattering volume $a_1$ and effective momentum $r_1$.
  The energy per particle $\bcE$ in the many-body system is calculated by resumming the ladder diagrams comprising both particle-particle and hole-hole intermediate states, \jo{following the novel advances recently developed by us
    in %J.M.~Alarc\'on and J.A.~Oller,
    Ann.Phys.{\bf 437},168741(2022). %\cite{Alarcon:2021kpx}.
  This allows to obtain  a renormalized result for $\bcE$ within generic cutoff regularization schemes, with $\bcE$ directly expressed in terms of the scattering parameters $a_1$ and $r_1$, \jo{once  the cut off is sent to infinity.}  
 The whole set of possible values of $a_1$ and $r_1$ is explored for the first time in the literature looking for minima in the energy per particle with $\bcE$ given as described.  
 They} are actually found, but a further inspection reveals that the associated scattering parameters give rise to resonance poles in the complex momentum-plane with positive imaginary part, which is at odds with the Hermiticity of the Hamiltonian.
  We also determine that these conflictive poles, with a pole-position momentum that is  smaller in absolute value than the Fermi momentum of the system, clearly impact the calculation of $\bcE$.
  As a result, we conclude that unpolarized spin-1/2 fermionic normal matter interacting in $P$-wave is not stable.
  We also study three universal parameters around the unitary limit.
  Finally, the whole set of values for the parameters $a_1$, $r_1$ is characterized according to whether they give rise to unallowed  poles and, if so, by attending to their pole positions relative to the Fermi momentum of the system explored. 
\end{abstract}
\maketitle

\section{Introduction}
  The stability of liquids, whether classical or quantum-mechanical, stems from the competition between kinetic energy (temperature), repulsive and attractive forces \cite{Hansen2013,Leggett2006}.
  Near the absolute zero, attractive forces allow the constituent particles to stay close to each other -- forming bound states -- while repulsion, usually at short distances \cite{Aziz1991}, avoids the collapse of the system in the thermodynamic limit.
  In the quantum case, it was recently shown theoretically \cite{Petrov2015}, and subsequently observed experimentally \cite{Cabrera2018,Cheiney2018,Semeghini2018}, that certain bosonic systems can stably liquify at low temperatures thanks to one of the competing interactions arising from quantum fluctuations.
  Fermions, on the other hand, obey the Pauli principle, which acts as a natural mechanism for short-distance repulsion \cite{fetter}. Therefore, attractive interactions on their own may stabilize fermionic quantum liquids.

  Over the past decade or so, there have been tremendous advances in the manipulation and characterization of low-energy atom-atom interactions in the $P$-wave channel \cite{Ticknor:2004,Schunck:2005,Zhang2004,Dong2016,Chang2020,Marcum2020}, their $S$-wave counterparts being well controlled and understood for some time now \cite{Chin2010}.
  The effective low-energy $P$-wave interactions can be tuned essentially at will, including near resonance, where it may become the dominant interaction in the system, and certainly does for spin-polarized systems.
  An important, yet difficult question due to the non-perturbative nature of the interaction, is whether fermions near a $P$-wave resonance can form a stable liquid. In Ref.~\cite{Gurarie:2007} a polarized degenerate  gas of identical fermions is studied near an isotropic or anisotropic Feshbach resonance \cite{Ticknor:2004}, and a controlled expansion is developed for the case of narrow resonances characterized by $|k_F/r_1|\ll 1$. 
  A much richer structure of phases is predicted \cite{Gurarie:2007} as compared with the case of $S$-wave interactions.  
  
  In this work, we consider instead the case of a spin-balanced, or unpolarized, degenerate Fermi gas interacting  in $P$-wave and address the question whether such system can be stable. Because of the Pauli exclusion principle the $P$-wave interactions of two fermions 
  take place necessarily in the triplet spin channel. For the $P$ wave in a spin-balanced Fermi system  to be more relevant at low energies than the $S$ wave  a suppression is required for the spin singlet interactions in which the $S$ wave takes place. It is beyond the scope of our research to determine which conditions could drive to this situation both in the laboratory or in Nature. Nonetheless, we consider that it is an interesting theoretical question to pursuit whether such system can bound.
  It is also worth noticing that $P$-wave interactions  provide important contributions to the equation of state of fermionic systems in nature  which naturally occur unpolarized, like neutron matter \cite{Holinde:1981se,Lacour:2009ej} that is of key importance in Astrophysics for the study of  neutron stars.
  Indeed, the interest in these objects has been boosted by the recent observation of gravitational waves emitted from the coalescence of a compact binary made by a black hole and a neutron star \cite{LIGOScientific:2021qlt}.

  In our study we describe  elastic fermion-fermion interaction in $P$-wave by the effective-range expansion (ERE) up to and including  the scattering volume, $a_1$, and the effective momentum, $r_1$. At least two low-energy scattering parameters are needed to end with a renormalized $P$-wave fermion-fermion interaction \cite{Alarcon:2021kpx,Bertulani:2002sz} \jo{within cutoff regularization when sending the cutoff to infinity}. 
  The many-body calculation of the energy density $\cE$ is based on the application of the formalism recently developed by \cite{Alarcon:2021kpx}, where a compact expression  is derived by resumming the ladder diagrams  for a spin-1/2 fermionic many-body system \jo{and, for the first time in the literature, providing renormalized results within a general cutoff regularization scheme after the cutoff is sent to infinity}.
The theory developed in Ref.~\cite{Alarcon:2021kpx} rests in the reformulation of many-body quantum field theory undertaken in Ref.~\cite{Oller:2001sn}, which determines the generating functional of Green functions in the fermionic many-body environment. The interested reader can also consult Ref.~\cite{Meissner:2001gz} for perturbative calculations up to next-to-leading order, and 
Ref.~\cite{Oller:2019ssq} for a recent review, considering also non-perturbative applications. We give here a brief account of the main features of this approach and refer to  Ref.~\cite{Alarcon:2021kpx} for a detailed exposition.

\jo{We would like to stress now that  resumming the two-body interactions with intermediate particle-particle and hole-hole states, $i.e.$ the ladder diagrams \cite{Thouless:1960anp}, is the leading interacting contribution for the calculation of the energy density $\cE$ as required by the  %dedicate some discussions to the corrections to our results for normal matter based on resumming the ladder diagrams in Fig.~\ref{fig.hartreefock} for arbitrary vacuum interactions.  For that we apply
  in-medium non-perturbative} power-counting of Ref.~\cite{Lacour:2009ej}, and  reviewed in Ref.~\cite{Oller:2019ssq}. %This power counting requires to resum the two-body interactions with intermediate particle-particle and hole-hole states as the leading interacting contribution for the calculation of the energy density $\cE$.
If the high-energy scale of the interactions is $\Lambda$ the expansion of in-medium diagrams is in powers of $k_F/\Lambda_\xi$ with $\Lambda_\xi\lesssim \Lambda$, so that subleading contributions suppressed by at least one extra factor of ${\cal O}(k_F/\Lambda_\xi)$. E.g. for the case of neutron matter, \jo{with pions included as explicit degrees of freedom in the theory}, the expansion scale is around $\sqrt{3}\pi f_\pi\simeq 0.5$~GeV  or $2.53$~fm$^{-1}$, corresponding to a density of 0.55~fm$^{-3}$ \cite{Oller:2019ssq}. \jo{ This is large enough to address the equations of state for pure neutron and symmetric neutral matter in the density regions of interest for many applications, like for studying  neutron stars \cite{Dobado:2011gd} or saturation of nuclear matter \cite{Lacour:2009ej}, respectively, among many others. See Ref.~\cite{Oller:2019ssq} for a recent review. The importance to achieve the resummation of the ladder diagrams for studying $\bcE$, as a starting point on top of which to build perturbative many-body corrections, has been stressed in many instances in the literature. Indeed, as originally explained by Thouless in his seminal paper \cite{Thouless:1960anp}, the Brueckner theory is a particular case of the ladder resummation when the intermediate states are only of the particle-particle type. For more recent papers along this direction, emphasizing an effective-field theory point of view, we can quote Refs.~\cite{Schafer:2005kg,Kaiser:2011cg,Kaiser:2012sr,Steele:2000qt,Lacour:2009ej,Hammer:2019poc,Boulet:2019wfd}. It also follows from these discussions and reasons  that the evaluation of $\bcE$ within the ladder resummation is certainly an interesting result. This calculation and its proper interpretation are the main aim of our present research. }

\jo{  The ladder resummation was finally performed algebraically for arbitrary vacuum fermion-fermion interactions in Ref.~\cite{Alarcon:2021kpx}. The case of  contact interactions was explicitly solved, providing renormalized results for a general cutoff regularization scheme. As commented above the approach of Ref.~\cite{Alarcon:2021kpx} is the basis for our present research, in which we extend the results given in the previous reference for a many-body system of spin 1/2 fermions that  interact in $P$ wave. This is so because we now} study the energy per particle $\bcE$ in the whole set of scattering parameters and the Fermi momentum $k_F$ of the system by employing two dimensionless parameters $u=\tanh(-1/a_1r_1^3)$ and
  $v=\tanh(k_F/r_1)$ with $u,\,v\in (-1,1)$.
  In the square $(-1,1)\otimes (-1,1)$ inside the $uv$-plane  
  we find four continuum one-dimensional sets of $u$ and $v$ values in which the zero-temperature equation of state exhibits minima, which would correspond to the equilibrium energy per particle. 
  Other properties of these minima, like the speed of sound and the compressibility coefficient, are calculated in App.~\ref{app.220317.1}.

  However, we further scrutinize the feasibility of such minima by studying the pole content of the  vacuum  $P$-wave scattering amplitude sharing the same $a_1$ and $r_1$ parameters.
  Unfortunately, we find that all these minima correspond to a vacuum pole structure which includes two resonant poles whose associated momenta have a  {\it positive} imaginary part.
  These configurations are excluded by general principles, since the momenta in the resonance pole positions should have negative imaginary part as required by the Hermiticity of the Hamiltonian \cite{gottfried.book}.

  Nonetheless, one should keep in mind that such unallowed poles could be artifacts of the ERE that lie beyond its radius of convergence. Then,  we have still explored the possibility that some of these conflicting poles, whose distance from the origin of momentum establishes  an upper bound for the radius of convergence of the ERE, could lie at  momenta  larger than $2k_F$.
  In this way, the presence of such forbidden poles does not exclude that the in-medium momenta involved in the calculation of $\bcE$  may be smaller than the ERE radius of convergence.  
  This has been done by evaluating the absolute value of the ratio between $k_F$ and the resonant momentum,  and it is always found that this ratio is larger than 1, so that these poles are relevant in the momentum range of the many-body system and clearly impact the calculation of $\bcE$.
  
  As a result of these considerations, we  conclude that it is not possible that a degenerate unpolarized fermionic system interacting in $P$ wave reaches a stable configuration in normal matter for sufficiently low density (such that one can apply the ERE for characterizing the scattering amplitude in vacuum).
  This conclusion is complementary to that of Ref.~\cite{Gurarie:2007} for a single-species polarized fermion gas, where it is obtained that the normal matter state is not stable.

  The structure of the manuscript is as follows. After this Introduction we discuss in Sec.~\ref{sec.220224.1b} the formalism employed to calculate the energy density $\cE$ and plot the energy per particle $\bcE$ for the whole set of scattering parameters $a_1$ and $r_1$ looking for its minima as a function of the Fermi momentum $k_F$. Section~\ref{sec.220224.1} is dedicated to discuss the pole structure of a $P$-wave partial-wave amplitude characterized by its scattering volume and effective momentum, and its implications for the feasibility of calculated $\bcE$, paying special attention to its minima. Conclusions are gathered in Sec.~\ref{sec.220225.1}. There is also an appendix dedicated to other properties of the minima in $\bcE$.
  
%%%%%%%%%%%%%%%%%%%%%%%%%%%%%%%%%%%%%%%%%%%%%%%%%%%%%%%%%%%%%%%%%%%%%%%%%%%%%%%%%
  \section{Calculation of the energy per particle}
  \label{sec.220224.1b}

\begin{figure}
\begin{center}
\begin{tabular}{c}
  \includegraphics[width=0.45\textwidth]{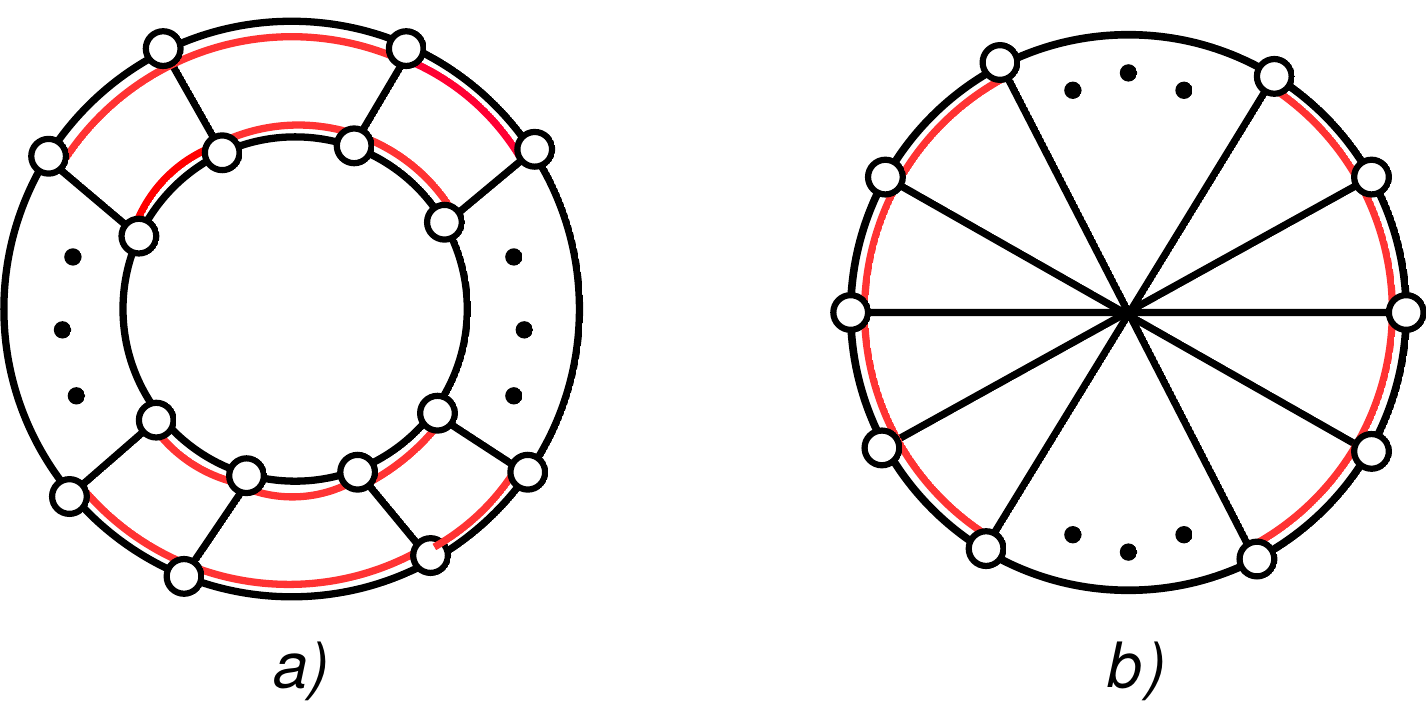}
\end{tabular}
\caption{{\small Resummation of the Hartree (a) and Fock (b) diagrams. The black and red lines running  in parallel are on-shell fermions with momenta less than $k_F$, and the $L_d$ loops are given by two sets of such lines facing each other (Hartree) or radially opposite (Fock).   The inwards single solid lines are $t_m$, and the dots indicate extra insertions of the interaction. \label{fig.hartreefock} }} 
\end{center}
\end{figure}

  Reference~\cite{Alarcon:2021kpx} performs the resummation of the Hartree and Fock diagrams in terms of an arbitrary vacuum $T$ matrix, called $t_V$, by resumming the ladder diagrams \cite{Thouless:1960anp}, shown in Fig.~\ref{fig.hartreefock}. 
  This resummation includes both particle-particle and hole-hole intermediate states interacting by insertions of $t_V$. We follow the notation of Ref.~\cite{Alarcon:2021kpx} and employ the resulting expression for the energy density, 
\begin{align}
\label{210325.1} 
\cE&=\frac{k_F^5}{10m\pi^2}-\frac{i}{2}{\rm Tr}
\ln\left(I-t_m L_d\right)\, .
\end{align}
Above, the first term on the right-hand side is the kinetic-energy density of a spin-$1/2$ Fermi gas, with fermion mass $m$ and Fermi momentum $k_F$.
 The second term is the interaction-energy density expressed in terms of a trace that is taken over all two-fermion states {\it inside} the Fermi sea, that is, with momenta under $k_F$ (this limitation is encoded in the loop function $L_d$ to be specified below). In Eq.~\eqref{210325.1}, $t_m$ is an on-shell two-fermion scattering amplitude that we define after some preliminaries are introduced.
 A detailed derivation of this formula is given Section~2.2 of Ref.~\cite{Alarcon:2021kpx}.

The expression for $L_d$ is 
\begin{align}
\label{190624.3}
L_d(p,\va)&\!=\!i\frac{m p}{16\pi^2}\sum_{\sigma_{1,2}}\!\int
\! d\hat{\vk}\,\theta({k_F}-|\va+p \hat{\vk}|)
\theta({k_F}-|\va-p\hat{\vk}|)\nn\\
&\times|p\hat{\vk}\sigma_1\sigma_2\rangle_A\,{_A\langle} p\hat{\vk}\sigma_1\sigma_2|\, .
\end{align}
We denote by $|\vp\sigma_1\sigma_2\ra_A$ an antisymmetric two-fermion state. The two fermions have spin 1/2, and third components of spin $\sigma_i$, $i=1,2$. Their relative momentum is $\vp\equiv (\vp_1-\vp_2)/2$, where $\vp_i$ is the momentum of the $i_{\rm th}$ particle, and the total momentum is $2\va\equiv \vp_1+\vp_2$. 
The Heaviside functions in Eq.~\eqref{190624.3} guarantee that the two fermions have momentum below $k_F$.  
The total energy of the on-shell pair is $2a^0=(\vp_1^2+\vp_2^2)/2m=(\va^2+\vp^2)/m$,
which fixes $p\equiv |\vp|$ for a given total momentum (since $\va$ is conserved in the scattering process).

The in-medium scattering amplitude $t_m(\va)$ stems from the iteration of $t_V$ with {\it mixed} two-fermion intermediate states in which one of them has momentum below $k_F$ while the momentum of the other one is unconstrained.
The loop function  $L_m(p,\va)$ associated to these intermediate states obeys the expression
\begin{align}
\label{190624.1}
&L_m(p,\va)\!=\!-\frac{m}{2}\!\sum_{\sigma_{1,2}}\!\int \! \frac{d\vk}{(2\pi)^3} \big[\theta({k_F}-|\va+\vk|) \\
  & + \theta({k_F}-|\va-\vk|)\big]
\frac{| \vk \sigma_1\sigma_2\rangle_A\, {_A\langle} \vk \sigma_1\sigma_2|}{\vk^2-\vp^2-i\ep}\, .\nn
\end{align}

The amplitude $t_m(\va)$ satisfies the following equation,
\begin{align}
\label{210706.1}
  t_m(\va)=t_V+t_V L_m(p,\va)t_m(\va)\, .
\end{align}
In Fig.~\ref{fig.hartreefock}, we draw the diagrams resummed by the interacting part of $\cE$ in Eq.~\eqref{210325.1}.
The Hartree diagrams are shown in Fig.~\ref{fig.hartreefock}(a) and the Fock ones in \ref{fig.hartreefock}(b).
The double lines (one in black and another in red running parallel to each other) are fermions inside the Fermi sea, while a single solid line corresponds to an in-medium interaction $t_m$. 
Notice that in the Hartree diagrams two facing sets of double lines made up a $L_d$,
while for the Fock diagrams this is so for radially opposite double lines.
The intermediate states in $L_d$ interact by $t_m$, and the three dots indicate further iterations of the product $t_m L_d$. 

The evaluation of the interaction energy is done using the partial-wave expansion developed in Ref.~\cite{Alarcon:2021kpx}.
The states in the partial-wave basis are denoted by $|J\mu \ell S p\ra$, with definite total angular momentum $J$, third component $\mu$, orbital angular momentum $\ell$ and total spin $S$.
By taking the trace in Eq.~\eqref{210325.1} in the plane-wave basis one has

\begin{align}
\label{200501.6}
{\cE}
&=\frac{k_F^5}{10m \pi^2}-\frac{2i}{m\pi^3}\!\!\sum_{J\mu \ell S}\!\!\chi(S\ell)^2
\!\! \int_0^{k_F} \!\!a^2da \int_0^{\sqrt{k_F^2-a^2}} \!\!\!\! pdp\\
&\times \la J\mu\ell S p|\ln\left(I-t_m(a\hvz) L_d(p,a\hvz)\right)|J\mu\ell S p\ra\, ,\nn
\end{align}
where rotational invariance allows to take $\va$ along the $z$ axis. The factor $\chi(S\ell)^2$ selects even values for the sum $S+\ell$ because of the Pauli exclusion principle, being 2 in that case. 

The integral equation satisfied by the partial-wave amplitudes (PWAs) in the many-body environment is more involved than in the vacuum case because of the extra mixing between PWAs due to the dependence of scattering on the total momenta. This extra mixing is accounted for within the formalism by the  matrix 
\begin{align}
\label{201228.4}
&{\cal B}_{J_2\mu \ell_2, J_1\mu \ell_1}=-2\chi(S\ell_2)\chi(S\ell_1)
\!\sum_{m_3 s_3} (m_3s_3\mu|\ell_2SJ_2)\\
&\times (m_3s_3\mu|\ell_1 S J_1)\! \int\! d\hvk \,Y_{\ell_2}^{m_3}(\hvk)^*Y_{\ell_1}^{m_3}(\hvk)
\theta({k_F}-|\vk-a\hvz|)\, .\nn
\end{align}
However, $\mu$ and $S$ are conserved in the scattering process \cite{Alarcon:2021kpx}. In the previous equation $Y_{\ell}^m(\hvk)$ is a spherical harmonic and $(m_1m_2m_3|j_1j_2j_3)$ is the Clebsh-Gordan coefficient for the addition of the angular momenta $\mathbf{j}_1+\mathbf{j}_2=\mathbf{j}_3$. 

For the case of contact interactions it is shown in Ref.~\cite{Alarcon:2021kpx} that it is possible to derive an algebraic expression for the vacuum off-shell PWAs ${t_{V}(p',p)}_{\alpha\beta}$ for $p$, $p'$ bounded (this is the case in the integral equation of Eq.~\eqref{210706.1}, where $p$ and $p'<2k_F$).
Using a cutoff regularization in a generic scheme these PWAs are renormalized by reproducing the effective-range expansion (ERE) up to some order, with the cutoff finally sent to infinity.
Within a matrix notation, so that the set of coupled PWAs is denoted by
the matrix ${\mathsf t_V}(p',p)$, one can show that 
\begin{align}
\label{210329.1}
&{\mathsf t_V}(p',p)=\frac{4\pi}{m}(p')^\ell \tau(p)(p)^\ell\,,\\
&(p)^\ell={\rm diag}({p^{}}^{\ell_1},\ldots,{p^{}}^{\ell_n})\,,  \nn
\end{align}
and analogously for $(p')^\ell$. In the previous equation, $\ell_i$ ($i=1,\ldots,n$)
is the orbital angular momentum of the $n$ coupled PWAs, and 
\begin{align}
\label{210329.2}
\tau(p)^{-1}\!=\!-(a)^{-1}\!+\!\frac{1}{2}(r)p^2\!+\!\sum_{i=2}^M(v^{(2i)}_\ell)p^{2i}\!-\!i (p^\ell)^{2}(p)\,,
\end{align}
with $M$ the order up to which the ERE is reproduced.   In this equation $(a)$, $(r)$ and $(v^{(2i)}_\ell)$ are $n\times n$ matrices corresponding to the scattering length, effective range and shape parameters, respectively.\footnote{Depending of the PWAs the dimension of the entries in $(a)$, $(r)$, etc, are different.} 
We note that a minimum value of $M$,  depending on the PWAs studied, is needed to achieve renormalizability. E.g. in the case of $P$-wave scattering, considered here, one needs at least $M= 2$, as also obtained in Ref.~\cite{Bertulani:2002sz}.

In terms of this expression, and using an analogous matrix notation for the set of PWAs in the many-body environment, 
\begin{align}
\label{201229.6}
&{\mathsf t_m}(p,p)=\frac{4\pi}{m}(p^\ell)
\left(\tau(p)^{-1}+[{\cal G}_m(p)]\right)^{-1}(p^\ell)\, ,\\
[{\cal G}_m(p)]&=-\frac{1}{\pi}\int_0^\infty\frac{k^2 dk}{k^2-p^2-i\ep}
(k^\ell)\cdot {\cal B}\cdot (k^\ell)\, .\nn
\end{align}
Notice that the integral over $k$ in $[{\cal G}_m(p)]$ is bounded because of the Heaviside function in Eq.~\eqref{201228.4} with $a\leq k_F$, cf. Eq.~\eqref{200501.6}, such that $k<2k_F$.

Now, we proceed to apply the previous formalism to a fermion liquid interacting via a spin-independent $P$-wave potential in momentum space
\begin{align}
  v(k,p)=\vk\cdot \vp(d_0+d_2(k^2+p^2))~,
  \end{align}
with the parameters $d_0$ and $d_2$ given in terms of the cutoff to reproduce the ERE up to and including the effective range. 
For this one-channel interaction the ERE simplifies since $(a)$, $(r)$, and $(v_1^{2i})$ in Eq.~\eqref{210329.2} are just numbers equal to $a_1$, $r_1$ and $v_1^{2i}$, respectively, as the scattering in vacuum is uncoupled. 
Had we used dimensional regularization, as in Refs.~\cite{Kaiser:2012sr,Kaiser:2011cg}, instead of cutoff regularization \cite{vanKolck:1998bw}, a renormalized result is possible reproducing only the scattering volume. However, as we argue in \cite{Alarcon:2021kpx}, we interpret it as an artifact of dimensional regularization in non-perturbative calculations because, as it is well known, the power-like divergences are set to zero in this particular regularization method.\footnote{See also Refs.~\cite{Phillips:1997xu,Oller:2017alp} for further discussions and more examples in vacuum scattering.} 

In the case of a $P$-wave interaction ($\ell=1$) of two identical fermions, Fermi statistics requires that $S=1$, i.e. the spin-triplet channel is selected, and for each $\mu$ only the PWAs with $J\geq |\mu|$ contribute.
For $\mu=0$ we have the mixing between the PWAs $^3P_0$ and $^3P_2$, while $^3P_1$ does not couple. In the case with $\mu=\pm 1$ the mixing is between $^3P_1$ and $^3P_2$, and for $\mu=\pm 2$ only the $^3P_2$ contributes. One can solve directly for ${\mathsf t_m}(p,p)$ in this case and in terms of it calculate the energy density $\cE$ applying Eq.~\eqref{210325.1}. The evaluation of the trace of the $\log$ is performed by diagonalizing the matrix in its argument which is possible because it is a unitary matrix \cite{Alarcon:2021kpx}. 

We study the energy per particle $\bar{\cE}=\cE/\rho$, where $\rho=k_F^3/3\pi^2$ is the number density, as a function of the scattering volume $a_1$ and the effective momentum $r_1$.
The dimensionless variables
\begin{align}
  x&=-1/a_1r_1^3~,\\
  y&=k_F/r_1 \nn
\end{align}
are introduced, so that
\begin{align}
\bar{\cE}=\frac{3k_F^2}{10m}f(x,y)=r_1^2 \frac{3y^2}{10m} f(x,y)~.
\end{align}
To look for the minima of $\bcE$ when varying $k_F$ for given values of $a_1$ and $r_1$ it is advantageous to consider the new coordinates
\begin{align}
  u&=\tanh(x)~,\\
  v&=\tanh(y)~,\nn
\end{align}
and then the whole $xy$-plane  is compressed to the finite extent $(-1,1)\otimes (-1,1)$ in the $uv$-plane.
 The minimum of $\bcE$ has to be searched along the $v$ axis, because when $k_F$ changes $u$ stays put. 
We look for them  numerically in steps of $10^{-3}$ in the $v$ variable for a given $u$.  
Numerical inaccuracy prevents us from taking smaller steps, which could {\it a priori} limit our search for potentially narrow minima.

\begin{figure}[ht]
\begin{center}
\begin{tabular}{c}
  \includegraphics[width=0.5\textwidth]{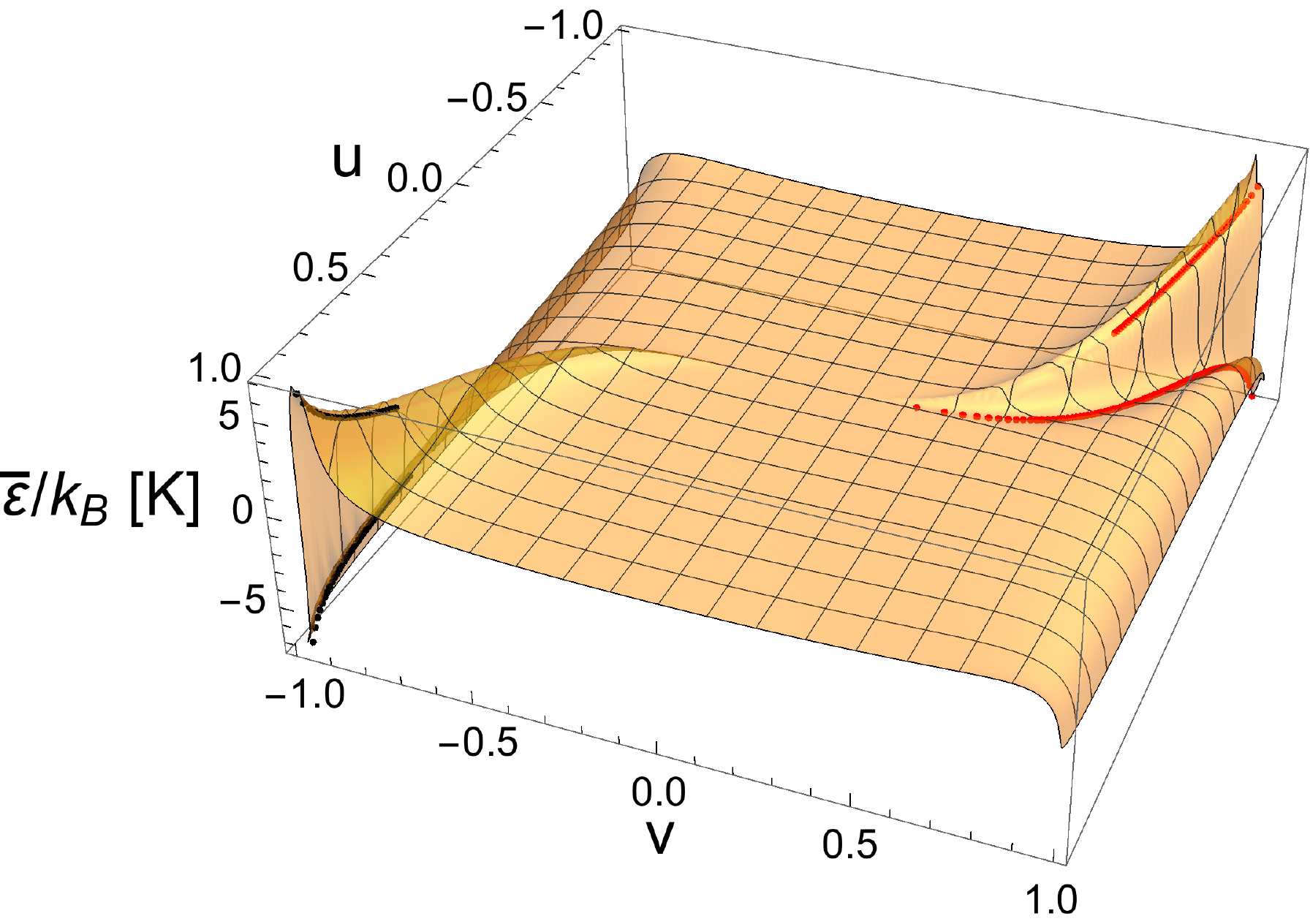}\\
\end{tabular}
\caption{{\small  The energy per particle $\bcE$ in units of kelvin is plotted  in the $uv$-plane.
  The positions of the minima are also indicated by the dots on top of the surface. 
 \label{fig.EN_minima_uv} }} 
\end{center}
\end{figure}

The energy per particle, in kelvin, $\bcE/k_B$, with $k_B$ the Boltzmann constant, is plotted in Fig.~\ref{fig.EN_minima_uv} in the $uv$-plane.  The filled dots in the same figure mark the positions of the minima.
 The value of  $\bcE$ in absolute terms results by taking, for concreteness, 
  $|r_1|=0.28\,a_B^{-1}$ \cite{Chang2020} and $m$ for $^6 \mathrm{Li}$ scattering, with $a_B$  the Bohr radius~\footnote{Within our conventions the effective range is multiplied by a  factor $-2$ compared to \cite{Chang2020}.}.     
Notice that for given values of $x$ and $y$ the sign of $r_1$ does not affect $\bcE$ since, as indicated above, $\bcE=r_1^2\frac{3y^2}{10 m}f(x,y)$.
In Fig.~\ref{fig.EN_minima_uv}, we observe four branches of minima within the whole set of parametric values for $a_1$ and $r_1$, two with $\bcE>0$ and the other two with $\bcE<0$.  
Among all of them, and within our numerical precision,  only one branch  can approach arbitrarily close to the unitary limit
at $u=v=0$ ({\it i.e.} $|a_1|,\,|r_1|\to \infty$, and $k_F\to 0$), though never strictly on it. 
This is the longest branch in the top right corner in Fig.~\ref{fig.EN_minima_uv}. 
Indeed, as we have checked explicitly,  the minimum for $u=0$ happens at  $k_F=0$.
 The other branches only occur for values of $|u|$ large enough that, for a given $r_1$,
implies a maximum value for $|a_1|$, and  stay away from the unitary limit.\footnote{We expect that temperature would not affect our results on $\bcE$ shown in Fig.~\ref{fig.EN_minima_uv} as long as it is much smaller than $\bcE$ expressed in kelvin degrees. This is the case in  ultracold fermionic atom experiments where temperature is typically around  a few tens of $\mu K$ (the only possible exception would be to stay almost on top the unitary limit where $\bcE$ also tends to vanish). The values of $\bcE$  along the minima can be better read from Figs.~\ref{fig.prop.enlt0} and \ref{fig.prop.engt0}.} Other properties of the minima of $\bcE$ are discussed in the Appendix~\ref{app.220317.1}. They are relegated there because the presence of these minima is very much affected by untenable pole structures associated to the ERE used for their calculation, as discussed in Sec.~\ref{sec.220224.1}.

Let us consider closely the interesting unitary limit with $x=0$ and work out the leading behavior in powers of $y$ of $f(0,y)=\bar{\cE}\frac{10m}{3k_F^2}$ as introduced above. To calculate $\cE$ we have the dimensionless combination $1-{\mathsf t_m}L_d$ inside a $\log$, cf. Eq.\eqref{210325.1}. In $k_F^3\tau(k_F)$ we can extract a global dimensionless factor $y^3=(k_F/r_1)^3$ and the rest adopts the form $\left(x+\frac{1}{2}y^2+v_1^{(2)}r_1 y^4-i y^3+{\cal O}(y^4)\right)^{-1}$, with $x=0$ in the unitary limit. We indicate the real and imaginary parts of ${\mathsf t_m}$ as ${\mathsf t_m^r}$, ${\mathsf t_m^i}$, respectively, and  $L_d$ is written as  $iL_d^i$. Therefore, ${\mathsf t^r_m}L_d^i$ 
is proportional to $y$ while $1-i{\mathsf t_m^i}L_d^i$ behaves as $1+{\cal O}(y^2)$, so that ${\rm arctan}\!\left({\mathsf t_m^r L_d^i}/(1+{\mathsf t^i_m}L_d^i)\right)\sim y$. We can then write $\bar{\cE}=\frac{3k_F^2}{10m}(1+\xi_1 y)$,  for $x=0$ and $|y|\ll 1$.  
Because of the non-perturbative contribution of the effective-range parameter the unitary limit depends on the order in which the limits $x\to 0$ and $y\to 0$ are taken. Following an analogous procedure as before it results that for $|y/x|\ll 1$ the function $f(x,y)$ behaves as $(1+\xi_2 y^3/x+\xi_3 y^5/x^2)$, where both $\xi_2$ and $\xi_3$ are universal since the higher-order shape parameters in the ERE, starting from $v_1^{(2)}$,  are ${\cal O}(y^7)$. 
Our calculation predicts the following values for these universal parameters
\begin{align}
\label{210419.1}
\xi_1&=-6.4054\,,~\xi_2=-0.9549\,,~\xi_3=0.2049\, .
\end{align}

At this point we would like to dedicate some discussions to the corrections to our results for normal matter based on resumming the ladder diagrams in Fig.~\ref{fig.hartreefock} for arbitrary vacuum interactions. 
For that we apply the power-counting developed in Ref.~\cite{Lacour:2009ej}, and more recently reviewed in Ref.~\cite{Oller:2019ssq}. This power counting requires to resum the two-body interactions with intermediate particle-particle and hole-hole states as the leading interacting contribution for the calculation of the energy density $\cE$. If the high-energy scale of the interactions is $\Lambda$ the expansion of in-medium diagrams is in powers of $k_F/\Lambda_\xi$ with $\Lambda_\xi\lesssim \Lambda$, so that subleading contributions suppressed by at least one extra factor of ${\cal O}(k_F/\Lambda_\xi)$. E.g. for the case of neutron matter the expansion scale is around $\sqrt{3}\pi f_\pi\simeq 0.5$~GeV  or $2.53$~fm$^{-1}$, corresponding to a density of 0.55~fm$^{-3}$ \cite{Oller:2019ssq}.
Therefore, according to the power counting of Ref.~\cite{Lacour:2009ej},  
$\xi_1$ and $\xi_2$ above should not receive further contributions from more complicated in-medium Feynman diagrams.
Thus, our study around the unitary limit of the energy per particle for small values of $k_F$ is particularly interesting and solid.

%%%%%%%%%%%%%%%%%%%%%%%%%%%%%%%%%%%%%%%%%%%%%%%%%%%%%%%%%%%%%%%%%%%%%%%%%%%%%%%%%
\section{Constraints from the pole structure in vacuum}
\label{sec.220224.1}

The secular equation for the poles of $\tau(p)$ in the uncoupled $P$-wave scattering reads
\begin{align}
  \label{220224.1}
  -\frac{1}{a_1}+\frac{1}{2}r_1 p^2-ip^3=0~.
\end{align}
This equation was already studied in the Appendix~E of our previous paper \cite{Alarcon:2021kpx} and applied for specific sets of values of the scattering parameters in $P$ wave.
Here we explore it for the full set of $a_1$ and $r_1$ parametric values, already considered in Fig.~\ref{fig.EN_minima_uv} for the calculation of $\bcE$.

Another important aspect not discussed in Ref.~\cite{Alarcon:2021kpx} is the possibility that {\it a priori} forbidden pole dispositions could drive to an acceptable calculation of $\bcE$. The reason is because the unallowed poles that stem from the use of the ERE  should occur at values of $p$ beyond the radius of convergence of the ERE.
So, if $k_F$ is small enough so that 2$k_F$ is smaller than the radius of convergence of the ERE,
one could still use that calculation for $\bcE$ even if the ERE generates unacceptable poles in the complex $p$-plane beyond its radius of convergence.
Notice that in the calculation of $t_m$, Eq.~\eqref{210706.1}, in-medium momenta $\vk$ up to $2k_F$ are involved in the function $L_m(p,a)$ of Eq.~\eqref{190624.1}. 

The zeroes in Eq.~\eqref{220224.1} are
\begin{align}
  \label{220129.1}
  p_1&=-\frac{ir_1}{6}\left(1+z+z^{-1}\right)~,\\
  p_2&=\frac{r_1}{6}\left(-i+e^{i\pi/6}z^{-1}-e^{-i\pi/6}z\right)~,\nn\\
  p_3&=\frac{r_1}{6}\left(-i-e^{-i\pi/6}z^{-1}+e^{i\pi/6}z\right)~,\nn
\end{align}
where,
\begin{align}
  \label{220129.6}
  \alpha_1&\equiv a_1^{1/3}=|a_1|^{1/3}{\rm sgn}(a_1)~,\\
  t&\equiv \frac{\alpha_1 r_1}{3\,2^{2/3}}=\frac{\alpha_1 r_1}{108^{1/3}}~,\nn\\
  z&\equiv \frac{t}{(1+t^3+\sqrt{1+2t^3})^{1/3}}~.\nn
\end{align}
From these definitions we have that  $z\in \mathbb{R}$, for $t\geq -1/2^{\frac{1}{3}}$. 
%the following properties for $z(t)$,  
%\begin{align}
%  \label{210129.8}
%&  z\in \mathbb{R}~,~\text{for }  t\geq -1/2^{\frac{1}{3}}~,\\
%&  z(-1/2^{\frac{1}{3}})=-1~,\nn\\
%&  z(+\infty)=1~,\nn\\
%&  z(-\infty)=e^{i2\pi/3}=(-1+i\sqrt{3})/2~.\nn 
%\end{align}

One can also work out straightforwardly the residues of the $P$-wave $S$ matrix at the different poles, which read 
\begin{align}
\label{220129.2}
\lim_{k\to p_i} (k-p_i) S_1(k)&=\lim_{k\to p_i}(k-p_i) \frac{2i k^3 }{-\frac{1}{a_1}+\frac{1}{2}r_1k^2-ik^3}\nn\\
&=\frac{-2p_i^3}{(p_i-p_k)(p_i-p_l)}~,
\end{align}
with $k,l\neq i$.  
It is well known since Heisenberg's papers on the $S$ matrix that the residue of $iS_\ell(k)$ at the bound-state poles must be positive.
There is a factor $i$ in front of the $S$ matrix because  $\jo{2\pi i {\rm{Res}}S_\ell(p_n)=}\left. \oint dk S_\ell(k)\right|_{p_i}=|C_i|^2$, integrating around the bound-state pole $p_i$ in a counterclockwise sense, where $C_i$ is a normalization constant multiplying the bound-state wave function  \cite{Hu:1948zz}. %Precisely this integration is $2\pi i {\rm{Res}}S_\ell(p_n)$ and h
\jo{Hence}, the residue of  $iS_\ell(k)$ must be positive at the bound-state pole. In the following we introduce the shorter notation ${\rm Res}_i$ to refer to the residue of $iS(k)$ at the $i_{\rm th}$ pole position, cf. Eq.~\eqref{220129.2}.\footnote{The requirement ${\rm Res}_i>0$ is equivalent to \jo{demand} that the residue of the $P$-wave scattering amplitude be positive, i.e. the coupling square $g^2$ of a bound state to the continuum must be positive. The relation is ${\rm Res}_i=g^2/2{\rm Im}p_i$, with ${\rm Im}p_i\geq 0$ for a bound-state pole.}

Our interest for the residue as a disclaimer for the acceptance or rejection of a bound-state pole  rests on  whether ${\rm Res}_i$ is positive or negative, respectively.
In this sense we can drop the factor $ip_i^3$ in ${\rm Res}_i$ because it is positive for a bound-state pole.\footnote{\jo{To see this let us denote by $i\lambda_i$, $\lambda_i\in \mathbb{R}$, the pole position for a bound or virtual state, then $ip_i^3=\lambda_i^3$, being positive for a bound state ($\lambda_i\geq 0$) and negative for a virtual state ($\lambda_i<0$). Therefore, the sign of ${\rm Res}_i$ is not affected by the factor $ip_i^3$ for a bound state, while its sign changes for a virtual state. As a result, keeping $ip_i^3$ from the numerator in Eq.~\eqref{220129.2}  has not effects whatsoever in our considerations that follow and we drop it.}} %1 Moved 
 Then, we introduce the quantity $\tau_i$ defined as
\begin{align}
\label{220129.4}
\tau_i=\frac{{\rm Res}_i}{ip_i^3}=\frac{-2}{(p_i-p_k)(p_i-p_l)}~,~k,l\neq i~,
\end{align}
as it follows from Eq.~\eqref{220129.2}.

A summary for  pole structure in the different ranges of values of $ t$ and $r_1$  is presented in  Table~\ref{tab.220130.1}.
Its derivation is discussed in detail in the Appendix~\ref{app.220618.1}.
It is found that the excluded regions have values for the scattering parameters such that $t r_1>0$, which is equivalent to having $a_1>0$, see Eq.~\eqref{220129.6}.

\begin{table}
  \begin{center}
    \begin{tabular}{|c|l|l|}
      \hline
      Case & $r_1<0$ & $r_1>0$ \\
      \hline
      $t<-1/2^{1/3}$ & $p_1$ v.s.           & b.s. PR \\
                    & $p_2$ b.s. PR        & v.s. \\
                    & $p_3$ b.s. {\bf NR}  & v.s. \\
      \hline
      $t=-1/2^{1/3}$ & $p_1=p_2$ {\bf Double b.s. pole} &  Double v.s. pole \\
                    & $p_3$ v.s.                       & b.s. PR \\
      \hline
      $-1/2^{1/3}<t<0$ & $p_1$ v.s. &  b.s. PR \\
                      & {\bf positive} ${\rm Im}p_{2,3}$ &  negative ${\rm Im}p_{2,3}$  \\ 
\hline
                $0<t$ & $p_1$ b.s. PR & v.s.  \\
               & negative ${\rm Im}p_{2,3}$ & {\bf positive}  ${\rm Im}p_{2,3}$ \\ 
\hline
    \end{tabular}
\caption{{\small  Pole structure for the different regions in $t$ and $r_1$. The values that are excluded are those with some boldface letter in them and happen for $t r_1>0$ or, equivalently, $a_1>0$. In the Table, PR(NR) is positive(negative) residue, and b.s.(v.s.) is bound(virtual) state pole.}
  \label{tab.220130.1}
}
    \end{center}
  \end{table}

%%%%%%%%%%%%%%%%%%%%%%%%%%%%%%%%%%%%%%%%%%%%%%%%%%%%%%%%%%%%%%%%%%%%%%%%%
\subsection{Checking the distance to the non allowed poles}
\label{sec.220130.3}

Inspection of Fig.~\ref{fig.EN_minima_uv} clearly reveals that the minima for $\bcE$
occur in the region of scattering parameters driving to a pole content in conflict with general principles, as summarized in  Table~\ref{tab.220130.1}. This is due to fact that in the minima $uv<0$,
or equivalently $tr_1>0$ because of the relations  
$ x=\arctanh u =-1/108 t^3$ and $r_1=k_F/\arctanh v$, with $k_F\geq 0$.
However, to directly rule out because of this reason the set of minima explored in Sec.~\ref{sec.220224.1b}, cf.~Fig.~\ref{fig.EN_minima_uv}, would not be justified.
The point is that the ERE has a radius of convergence given by the position of the nearest singularity to threshold, so that out of this convergence region it could give rise to artifacts, like the presence of unacceptable poles.
However, within its radius of convergence it is perfectly legitimate to keep using the ERE.
This reasoning was not taken into account in our recent paper \cite{Alarcon:2021kpx}, which directly rejected the results for $\bcE$ when using the ERE with parameters given rise to unacceptable poles.
Therefore, a deeper analysis on the use of the ERE is also needed to fully appreciate the range of applicability of the results  in Ref.~\cite{Alarcon:2021kpx}. 

\begin{table}
  \begin{center}
    \begin{tabular}{|r|r|r|r|r|r|}
      \hline
 ${^{2S+1}L_J}$&   $a_1$ [fm$^3$] & $r_1$ [fm$^{-1}$] & $t$ & $p_3$ [fm$^{-1}$] & $\tau_3$ [fm$^2$]\\
      \hline
      ${^1P}_1$ & 2.759 & $-6.54$ & $-1.926$ & $i\,0.352$ & $-1.035$ \\
      ${^3P}_1$ & 1.536 & $-8.40$ & $-2.059$ & $i\,0.412$ & $-0.668$ \\
      \hline
  \end{tabular}
\caption{ {\small Uncoupled nucleon-nucleon $P$-wave PWAs with positive $a_1$, giving rise to excluded pole structures when considering the ERE up to and including $a_1$ and $r_1$. We give $a_1$, $r_1$, $t<-1/2^{1/3}$, the conflictive bound-state pole $p_3$ and its residue $\tau_3$ for each PWA.} \label{tab.220226.1}}
  \end{center}
  \end{table}
As illustration of these points let us consider the application of the ERE in nucleon-nucleon scattering,  where one can find uncoupled $P$-wave PWAs with positive scattering volume,
like  the ${^1P}_1$ and ${^3P}_1$ ones.\footnote{We use the standard notation ${^{2S+1}L}_J$ to denote  PWAs.}
The central values of $a_1$ and $r_1$  for these $P$ waves,  taken from Ref.~\cite{Perez:2014waa}, are shown in Table~\ref{tab.220226.1}.
In the same table we also give the resulting values of $t$ (which are smaller than $-1/2^{1/3}$), the problematic pole position $p_3$, Eq.~\eqref{210130.1},  and the unacceptable $\tau_3<0$, Eq.~\eqref{210130.2}.
In the case of nucleon-nucleon scattering the radius of convergence of the ERE is  $|p|<m_\pi/2\approx 0.35$~fm$^{-1}$, which signals the onset of the left-hand cut due to one-pion exchange for $p^2<-m_\pi^2/4$, with $m_\pi$ the pion mass. 
We observe that for the ${^1P}_1$ partial-wave amplitude the value of $|p_3|$ is almost coincident with this upper limit. 
For the ${^3P}_1$ PWA the value of $|p_3|$ is a bit larger, around 0.41~fm$^{-1}$.
We then learn from these examples extracted from nucleon-nucleon scattering that {\it physical} values of $a_1$ and $r_1$ can drive to  non-allowed pole structures, with the conflicting poles happening (barely) out of the convergence region of the ERE. They cannot lie inside the radius of convergence for {\it finite-order} ERE, if this expansion is reasonable and well-behaved in its application.\footnote{\jo{This is perfectly the case for the $^1P_1$ and $^3P_1$ nucleon-nucleon PWAs since higher shape parameters in the ERE are given in Ref.~\cite{Perez:2014waa}, up to $v_4$, and we have checked that the unallowed bound state for each PWA remains stable if higher-order terms in the ERE are considered.}}

Contrarily to the case of nucleon-nucleon scattering we do not know the radius of convergence of the ERE for the results of $\bcE$ shown in Fig.~\ref{fig.EN_minima_uv}.
However, for each set of ERE parameters we can determine the position of the unacceptable poles for $a_1>0$ in the complex $p$-plane.
Attending to Table~\ref{tab.220130.1}, \jo{it turns out that in the regions of the $uv$-plane where the minima lie} the modulus of these non-allowed poles positions  is $|p_2|$.  
For our next considerations it is important to realize that $|p_2|$ is larger or equal than the radius of convergence of the ERE (accepting that it is well behaved with a progressive improvement in the results as the order is increased). \jo{That is, $|p_2|$ corresponding to an unallowed pole is an upper bound for the radius of convergence of the ERE.} 
We then compare it to $k_F$, and  if $2k_F> |p_2|$  then the results obtained for $\bcE$ are calculated by employing the ERE beyond its convergence radius, cf. Eq.~\eqref{201229.6}. 

In this regard, let us notice  that a resonant pole $p_R$ in the complex momentum-space with positive imaginary part also implies a positive imaginary part of the pole position in energy, $E_R\equiv M_R+i\Gamma/2$, with $M_R$ and $\Gamma$ real positive numbers.
   In the $P$-wave scattering amplitude this pole gives rise to a term of the form $-g^2/(E-M_R-i\Gamma/2)$ that produces phase shifts according to the argument of $M_R-E-i\Gamma/2$. Therefore, the phase shifts are negative and cross $-90$ degrees at the resonance mass. However, a proper resonance behavior must give rise to positive phase shifts because of the delay in the phase  of the scattering state trapped  by the pseudo-stationary resonance.

%%%%%%%%%%%%%%%%%%%%%%%%%%%%%%%%%%%%%%%%%%%%%%%%%%%%%%%%%%%%%%
   \subsubsection{$u<0$ and $v>0$}
   \label{sec.220316.1}

Let us first consider  the region $u<0$ ($t>0$) and $v>0$ in Fig.~\ref{fig.EN_minima_uv}. According to Table~\ref{tab.220130.1} the problem arises from the fact that ${\rm Im}p_{2,3}$ is positive, so that we evaluate 
 the ratio  $k_F/|p_2|$ as a function $t$ along the two curves where the minima lie. The procedure is the following: Given $t$  we can calculate along each curve $u(t)$ and $v(t)$  so that the sought ratio, cf. Eq.~\eqref{220129.1},  is 
\begin{align}
  \label{210130.9}
  \frac{k_F}{|p_2(t)|}&=\frac{6y(t)}{\left|-i+e^{i\pi/6}u(t)^{-1}-e^{-i\pi/6}u(t)\right|}~.
\end{align}
If this ratio were much smaller than 1 then we could argue that the poles $p_2$ and $p_3$ are not directly relevant for the densities involved in the many-body system, not providing resonant contributions. Conversely, if this ratio were  larger than 1 then the unphysical resonance phenomenon would be relevant and the associated minimum should be excluded.

Indeed, let us notice that from Eq.~\eqref{210130.5} it follows that
\begin{align}
\frac{2{\rm Re} p_2-{\rm Im} p_2}{r_1}=\frac{1-z}{12z}\left(2\sqrt{3}-1+(2\sqrt{3}+1)z\right)>0
\end{align}
since $0<z<1$. Therefore, the Laurent series in momentum of the $P$-wave amplitude provided by the ERE around the resonance pole $p_{2}$ or $p_3$, selecting the one with positive real part, converges along the physical real axis for 
\begin{align}
  \label{220227.2}
  0<p<|{\rm Re}p_2|\left(1+2\sqrt{1-\left( \frac{{\rm Im}p_2}{2{\rm Re} p_2}\right)^2}\right)~.
  \end{align}
%}$,
Thus, one has resonant contributions from this pole affecting the PWA for physical values of $p$.

To be more definitive let us show the typical behavior of the $P$-wave partial-wave amplitude for the minima with $uv<0$
in Fig.~\ref{fig.EN_minima_uv}. 
The phase shift, $\delta$, and the modulus squared of the resulting $P$-wave scattering amplitudes, $(\sin^2\delta)/p^2$,
are shown in the left and right panels of Fig.~\ref{fig.220316.1} (we take units such that $r_1=1$), respectively.
There, the top row is for the minimum with $u=-0.2$, $v=0.67$ (belonging to the branch having $\bcE<0$),
and the bottom row corresponds to the minimum with $u=-0.50$, $v=0.71$ (with $\bcE>0$).
 The values for $a_1$ and $k_F$ in the minima selected are:  $a_1=4.93~r_1^{-3}$ and $k_F=0.79~r_1$ for $u=-0.2,\,v=0.67$;  $a_1=1.82~r_1^3$ and $k_F=0.90~r_1$ for $u=-0.50,\,v=0.71$.
We can observe clearly marked resonant behaviors, with strong peaks in the square of the amplitudes and rapid varying phase shifts happening in both cases for $p<k_F$.
This unacceptable strong resonance signal should clearly affect the calculation of $\bcE$,
which then becomes non trustable and it should be rejected. 
Notice that these phase shifts have an inverse resonance shape because of the reason explained above associated with the wrong sign of the imaginary part of the momentum in the pole position.
Both minima drive to qualitatively similar behaviors of the $P$-wave scattering amplitude.

%%%%%%%%%%%%%%%%%%%%%%%%%%%%%%%%%%%%%%%%%%%%%%%%%%%%%%%%%
{\bf Region around the unitary limit:} 
One has then that  $u\to 0^-$, $t\to +\infty$, $z(t)\to 1$, and 
\begin{align}
  \label{210130.7}
p_2=r_1\left(\frac{1}{3\sqrt{6}t^{3/2}}+i\frac{1}{54t^3}\right)+{\cal O}(t^{-9/2})~.
\end{align}
It is not possible to get rid of this pole in the unitary limit since
\begin{align}
  \label{210130.8}
\frac{k_F}{|p_2|}\to  3\sqrt{6} yt^{3/2}~,
\end{align}
and it diverges for $t\to +\infty$. 

%%%%%%%%%%%%%%%%%%%%%%%%%%%%%%%%%%%%%%%%%%%%%%%%
\subsubsection{$u>0$ and $v<0$}
\label{sec.220316.2}

We can also proceed similarly for the two curves of minima in the region with $u>0$ ($t<0)$ and $v<0$.
We distinguish two further regions:

The region $-1/2^{1/3}<t<0$, $r_1<0$,  is excluded according to Table~\ref{tab.220130.1}. The criterion %for the possibility of reconsidering or definitely excluding this region is the
to be studied is the same as already expressed in Eq.~\eqref{210130.9}. 
 In this region $-1<z<0$, and within it along $-1<z<-(2\sqrt{3}-1)/(2\sqrt{3}+1)$ the difference $2|{\rm Re}p_2|-|{\rm Im}p_2|<0$. 
 However, this requires that $0.0185<u<0.055$ where there are no minima according to Fig.~\ref{fig.EN_minima_uv}.
 Therefore, the Laurent series around the resonance pole still converges within the interval  in Eq.~\eqref{220227.2} for the minima in the lower left quadrant of the $uv$-plane in Fig.~\ref{fig.EN_minima_uv}. 

\begin{figure*}[ht]
\begin{center}
\begin{tabular}{cc}
  \includegraphics[width=0.45\textwidth]{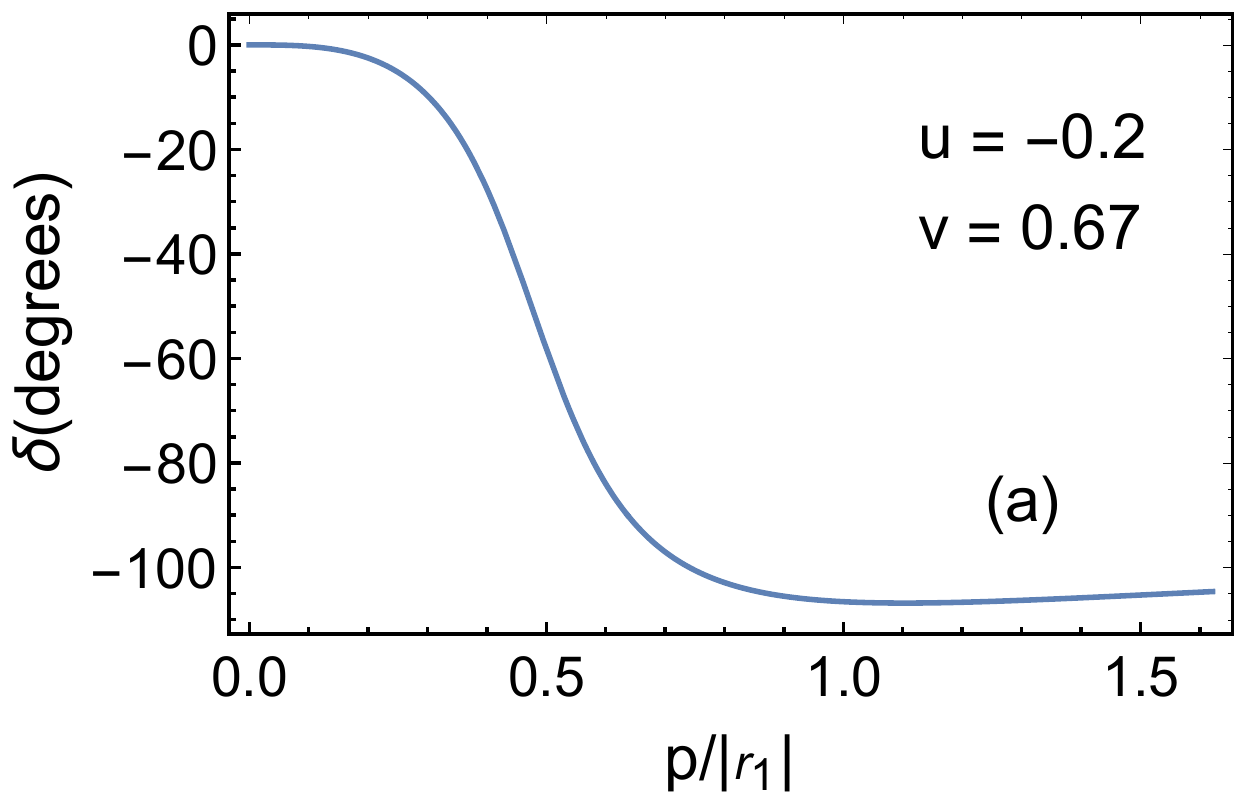} &  \includegraphics[width=0.45\textwidth]{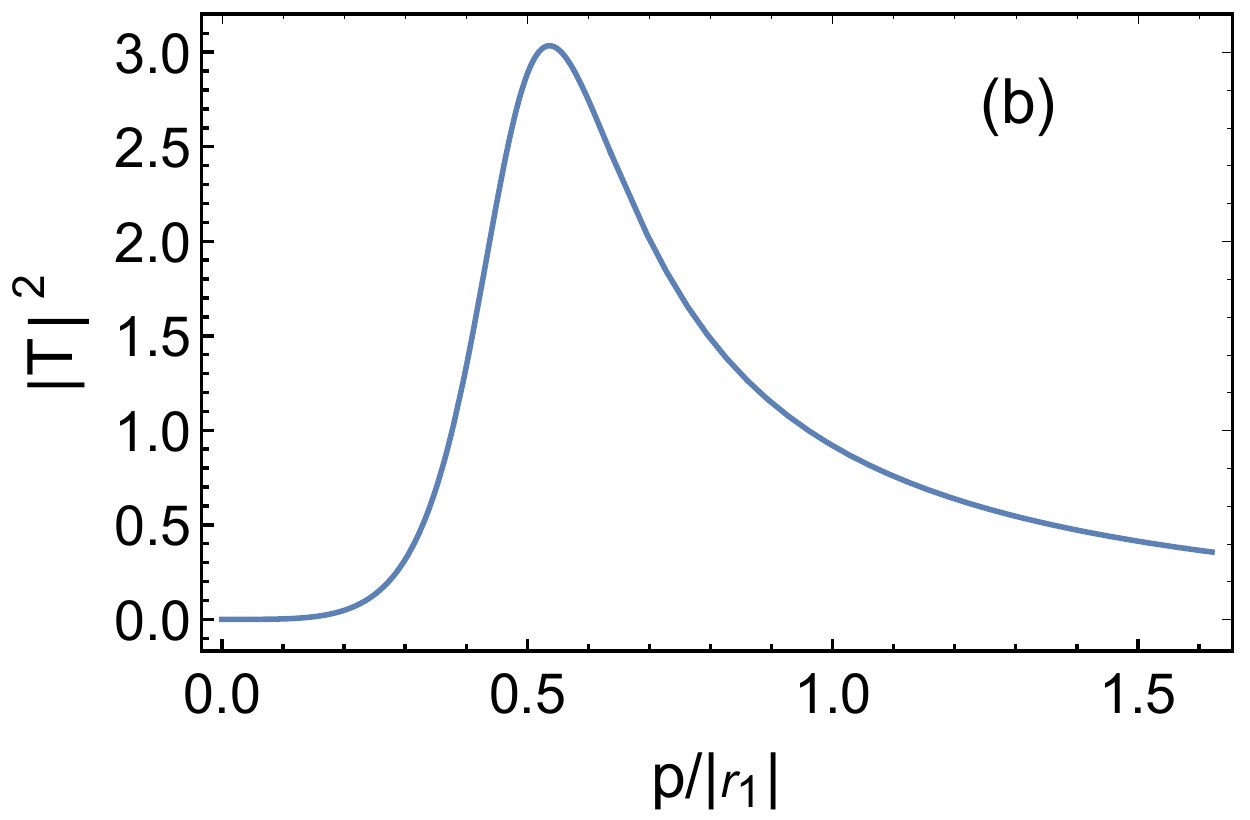}\\
  \includegraphics[width=0.45\textwidth]{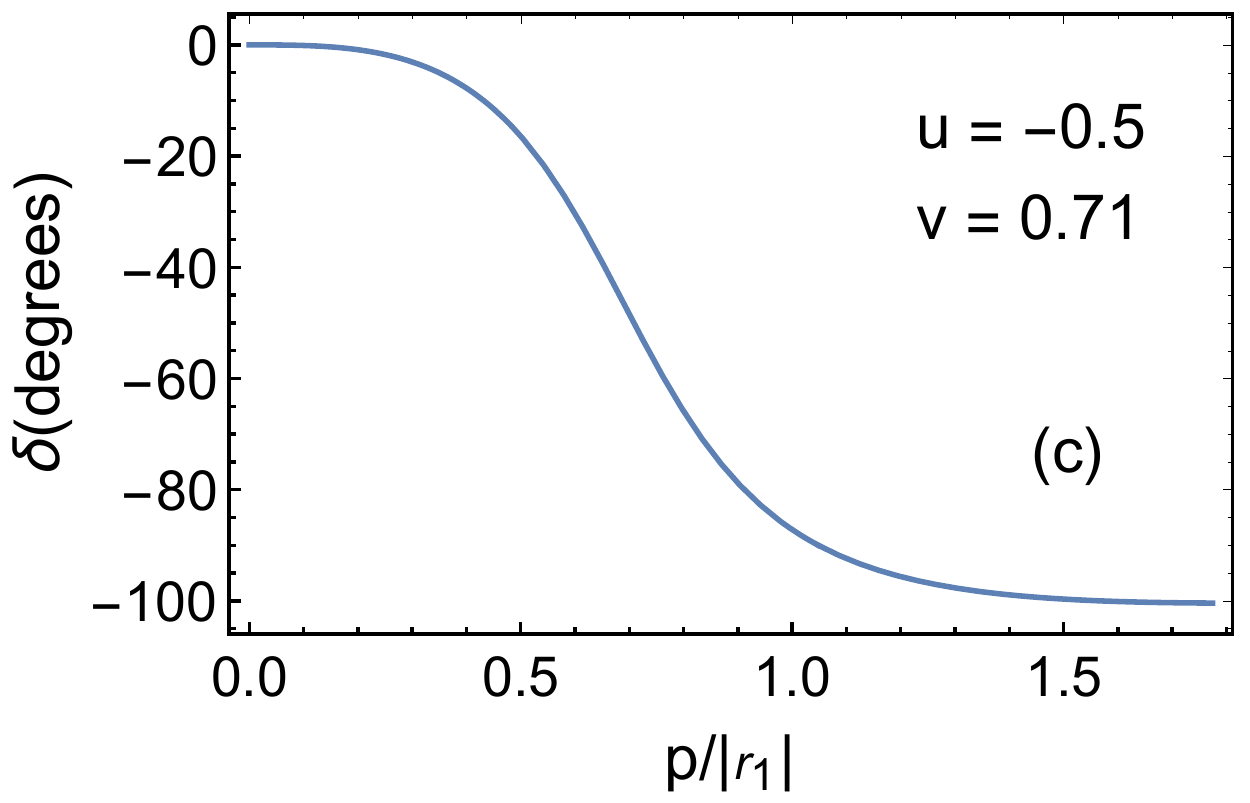} &  \includegraphics[width=0.45\textwidth]{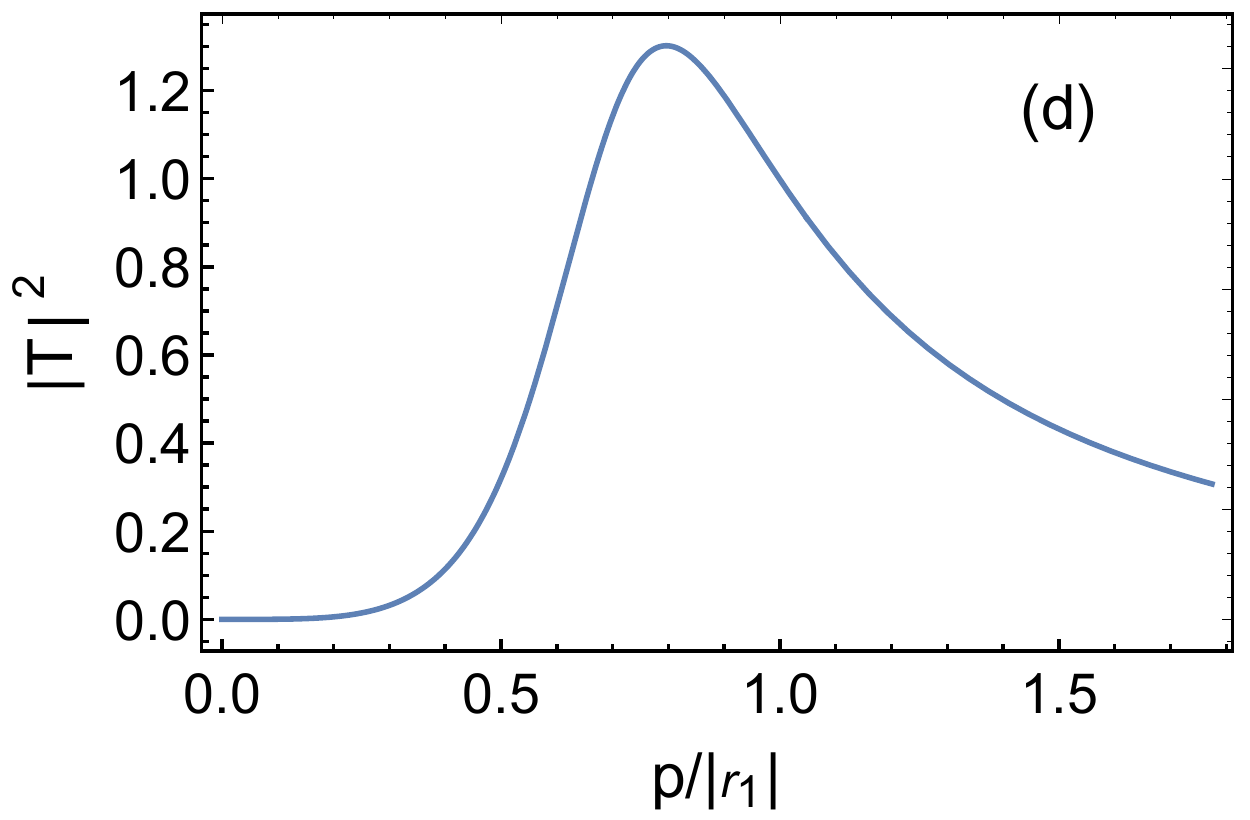}
\end{tabular}
\caption{{\small  Phase shifts and modulus squared of the $P$-wave scattering amplitude as a function of momentum in units of $r_1$ for two minima with $u<0$ and $v>0$. 
  The upper two panels correspond to the minimum $u=-0.20,\, v=0.67$ with $\bcE<0$, and the lower panels are for the  minimum $u=-0.50,\,v=0.71$ with $\bcE>0$.
    The interval of values shown in the $p$ axis runs from 0 up to $2k_F$.}
 \label{fig.220316.1}} 
\end{center}
\end{figure*}

\begin{figure*}[ht]
\begin{center}
\begin{tabular}{cc}
  \includegraphics[width=0.45\textwidth]{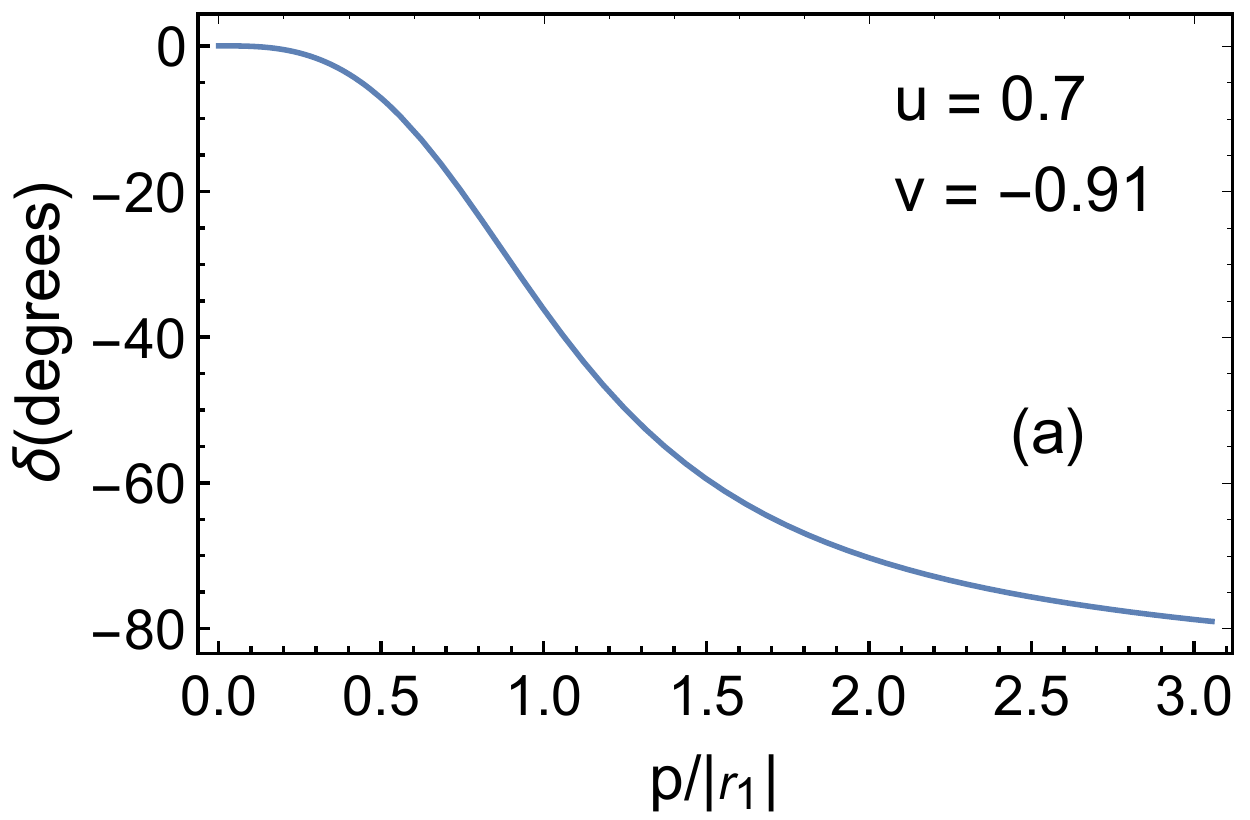} &  \includegraphics[width=0.45\textwidth]{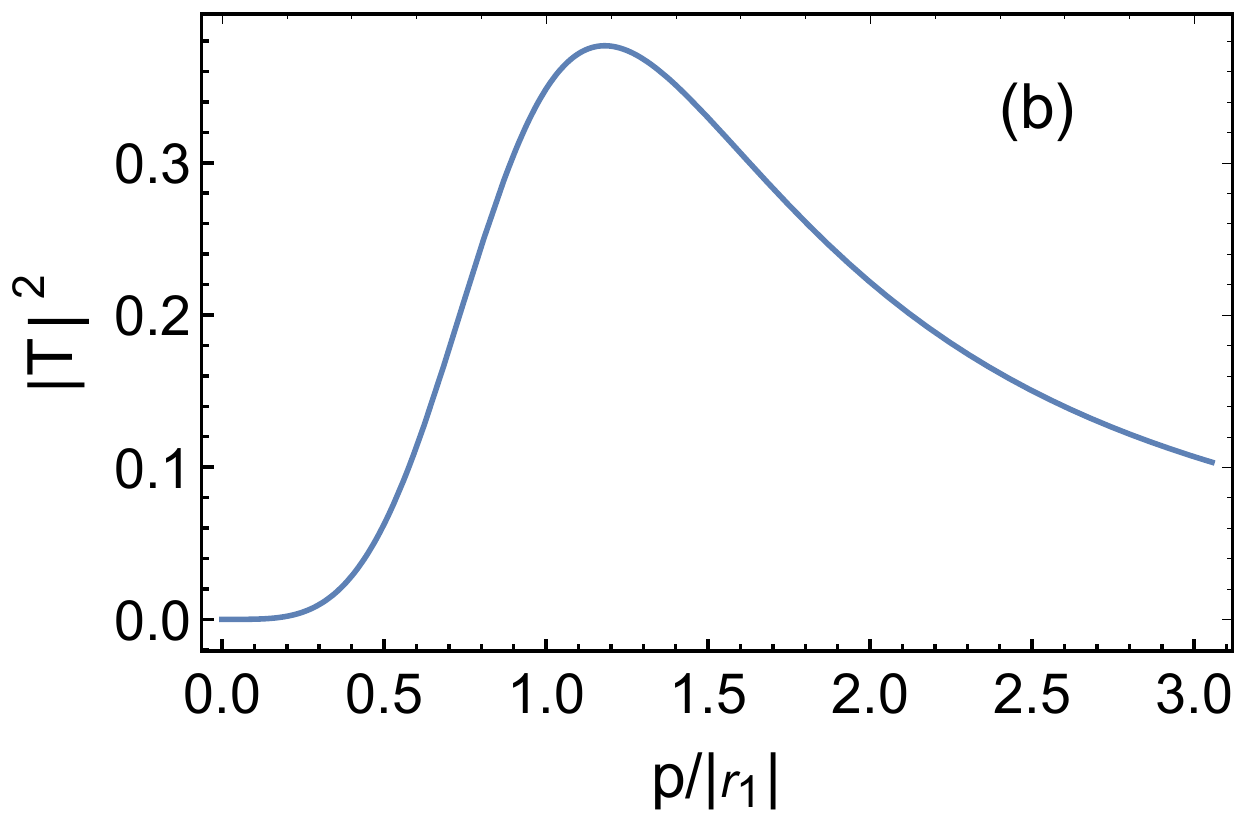}\\
  \includegraphics[width=0.45\textwidth]{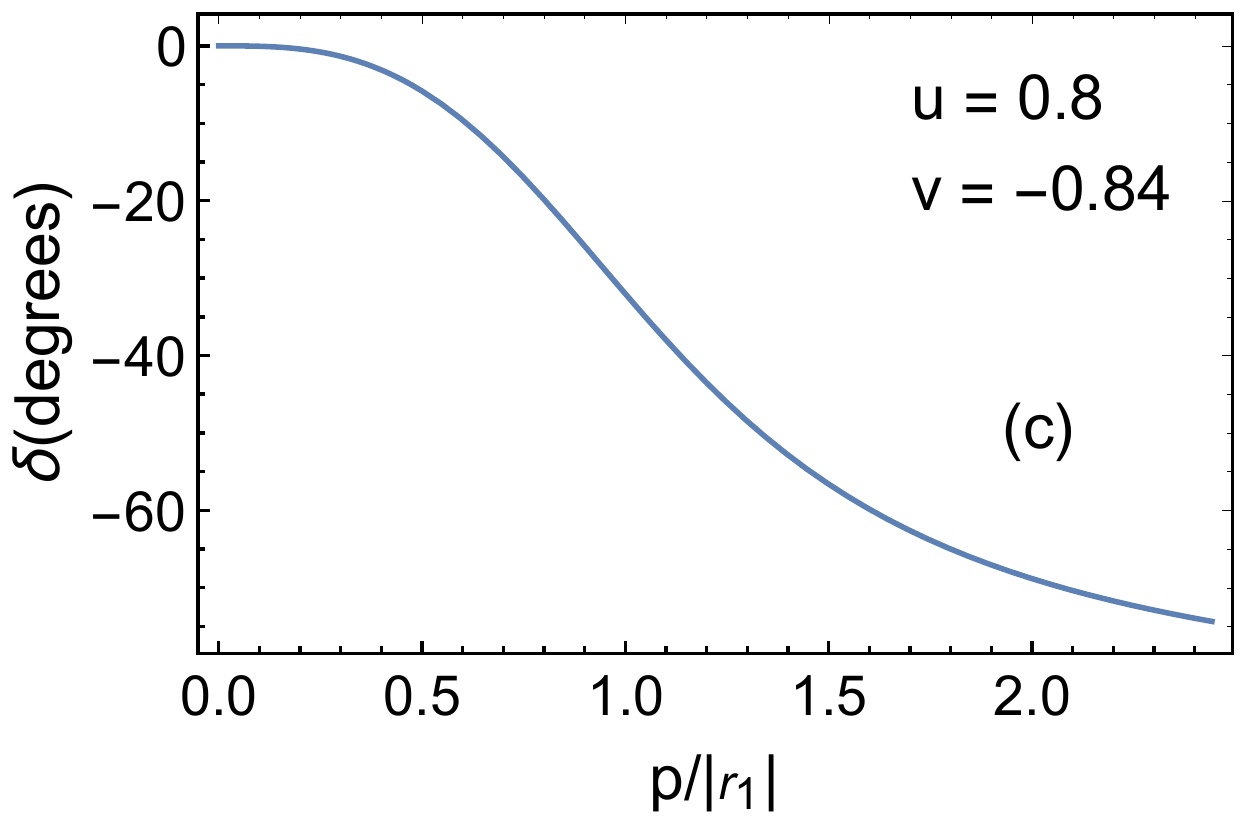} &  \includegraphics[width=0.45\textwidth]{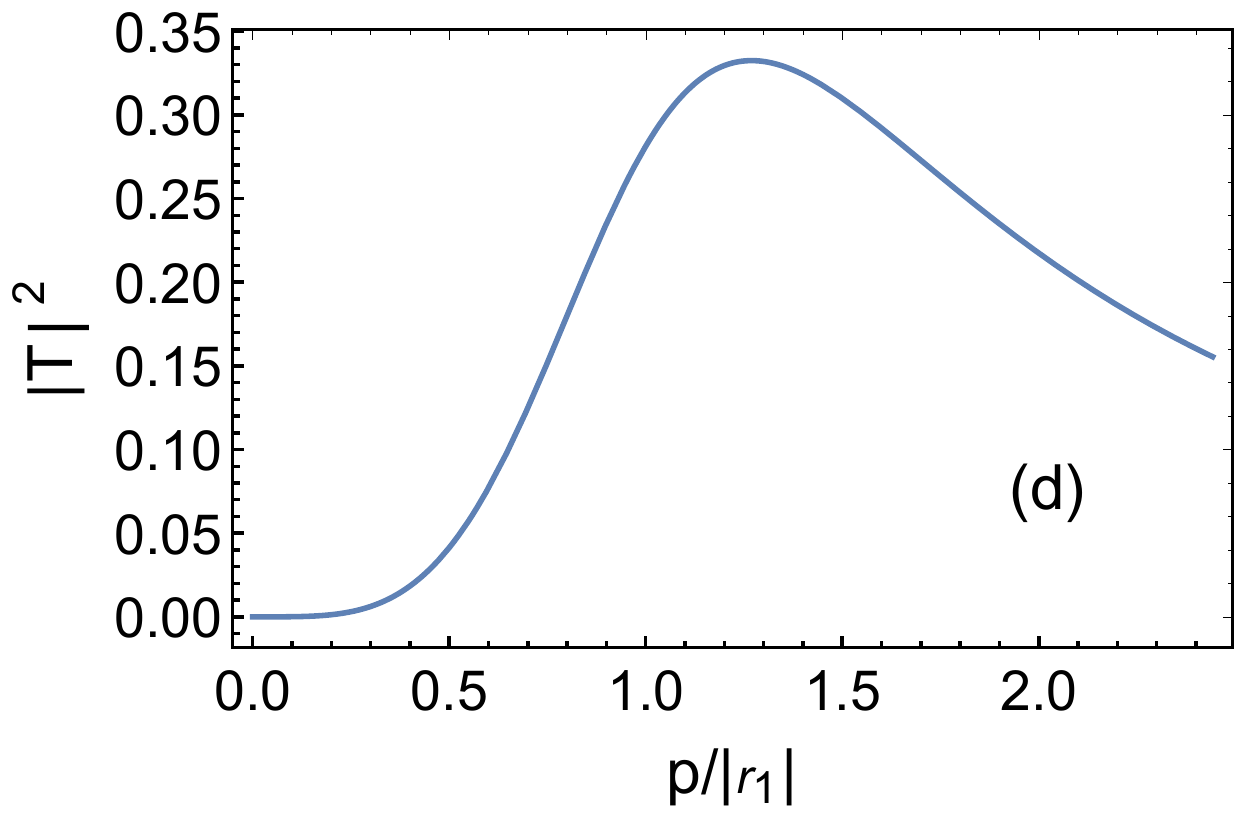}
\end{tabular}
\caption{{\small  Phase shifts and modulus squared of the $P$-wave scattering amplitude as a function of momentum 
in units of $|r_1|$ for two minima with $u>0$ and $v<0$. 
   The upper two panels correspond to the minimum $u=0.70,\,v=-0.91$ with $\bcE<0$, and the lower ones are for the minimum $u=0.80,\,v=-0.84$ with $\bcE>0$. 
    The interval of values shown in the $p$ axis runs from 0 up to $2k_F$.}
 \label{fig.220316.2} } 
\end{center}
\end{figure*}

 Analogously as for the case above with $u<0$, $v>0$, let us show the typical behavior of the $P$-wave scattering amplitude that is driven by the presence of these unacceptable resonance poles in the complex $p$-plane with positive imaginary part.
 We take two typical minima with $u=0.70$, $v=-0.91$ in the lower branch (with $\bcE<0$),
 and $u=0.80$, $v=-0.84$ in the upper branch (with $\bcE>0$).
 The resulting scattering volume $a_1$ and Fermi momentum $k_F$ are:
 $a_1=1.15~|r_1|^{-3}$ and $k_F=1.54~|r_1|$ for the minimum with $\bcE<0$;  
 $a_1=0.91~|r_1|^{-3}$ and $k_F=1.23~|r_1|$ for the minimum with $\bcE>0$.
  The  phase shifts and amplitude squared that follow are shown in the
 top and bottom rows in  Fig.~\ref{fig.220316.2} for the lower- and upper-branch minima, respectively.
 In the figure we take units such that $r_1=-1$. We observe again a clear resonance shape in the phase shifts and modulus
 squared of the $P$-wave partial-wave amplitude, though now the resonance is wider as compared with the case in Fig.~\ref{fig.220316.2}. 
 This is due to the larger imaginary part of $p_2$ for the poles with $u>0$ than for those with $u<0$.
 For instance, for the lower-branch minimum with $u<0$ in Fig.~\ref{fig.220316.1} we have $p_2=0.48+i\,0.15~r_1$, while for the analogous one with $u>0$ in Fig.~\ref{fig.220316.2} the pole position is $p_2=0.80+i\,0.66~|r_1|$. 
 The peak positions in Fig.~\ref{fig.220316.2} are around $k_F$ in both cases and hence the calculations of $\bcE$ are affected by the presence of this wrong strong resonance signals and should be discarded.

For the other  region with $t<-1/2^{1/3}$ the problem according to Table~\ref{tab.220130.1} is that the bound-state pole $p_3$ has a negative residue.  
The quotient is now
\begin{align}
  \label{210130.8}
  \frac{k_F}{|p_3(t)|}&=\frac{6|y(t)|}{\left|-i-e^{-i\pi/6}u(t)^{-1}+e^{i\pi/6}u(t)\right|}\\
  &=\frac{6|y(t)|}{1-\cos\phi-\sqrt{3}\sin\phi}~.\nn
%  \ll 1 ~. \nn
  \end{align}

However, this region is not relevant for the minima in Fig.~\ref{fig.EN_minima_uv} because
$t\leq -1/2^{1/3}$ implies that $u\leq 0.01852$, which is out of the region with minima, as one can check  by simple inspection of Fig.~\ref{fig.EN_minima_uv}.

%%%%%%%%%%%%%%%%%%%%%%%%%%%%%%%%%%%%%%%%%%%%%%%%%%%%%%%
\subsection{Discussion}
\label{sec.220226.1}

\begin{figure*}[ht]
\begin{center}
\begin{tabular}{c}
  \includegraphics[width=0.45\textwidth]{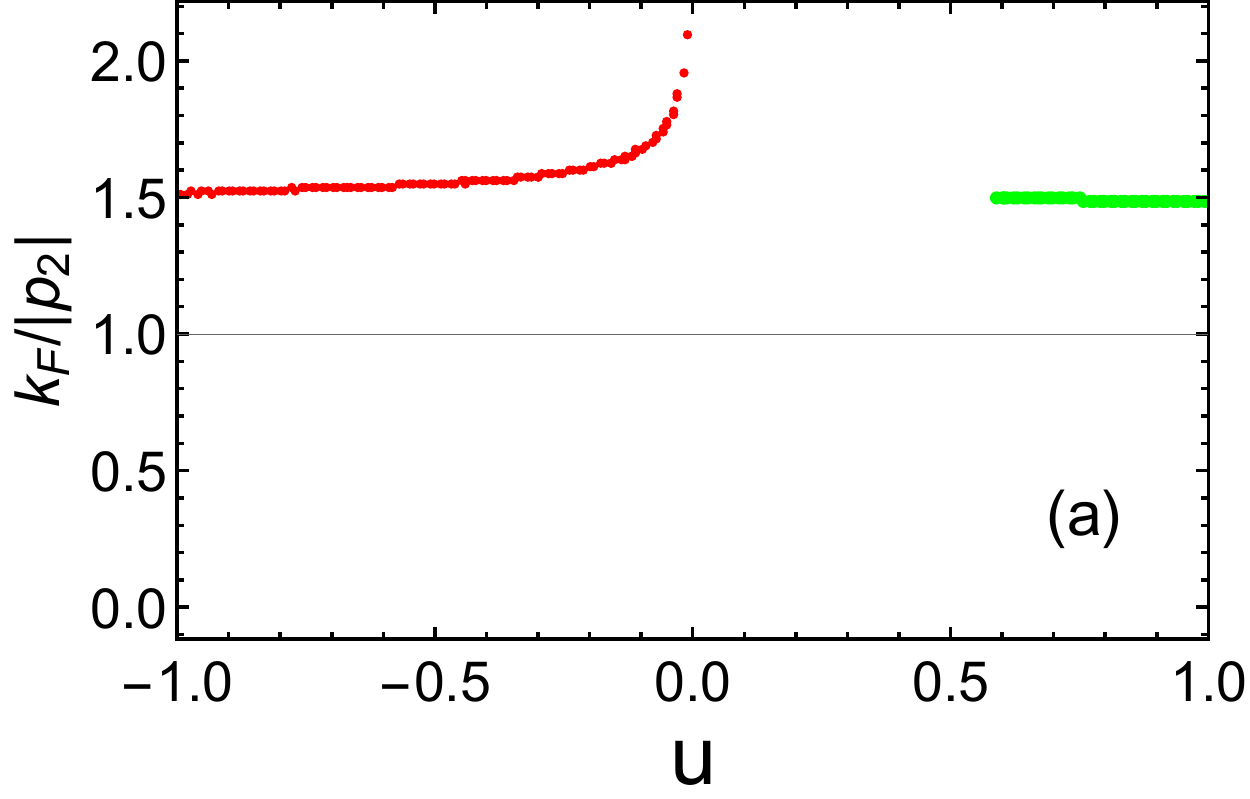} \hspace{1cm}  \includegraphics[width=0.45\textwidth]{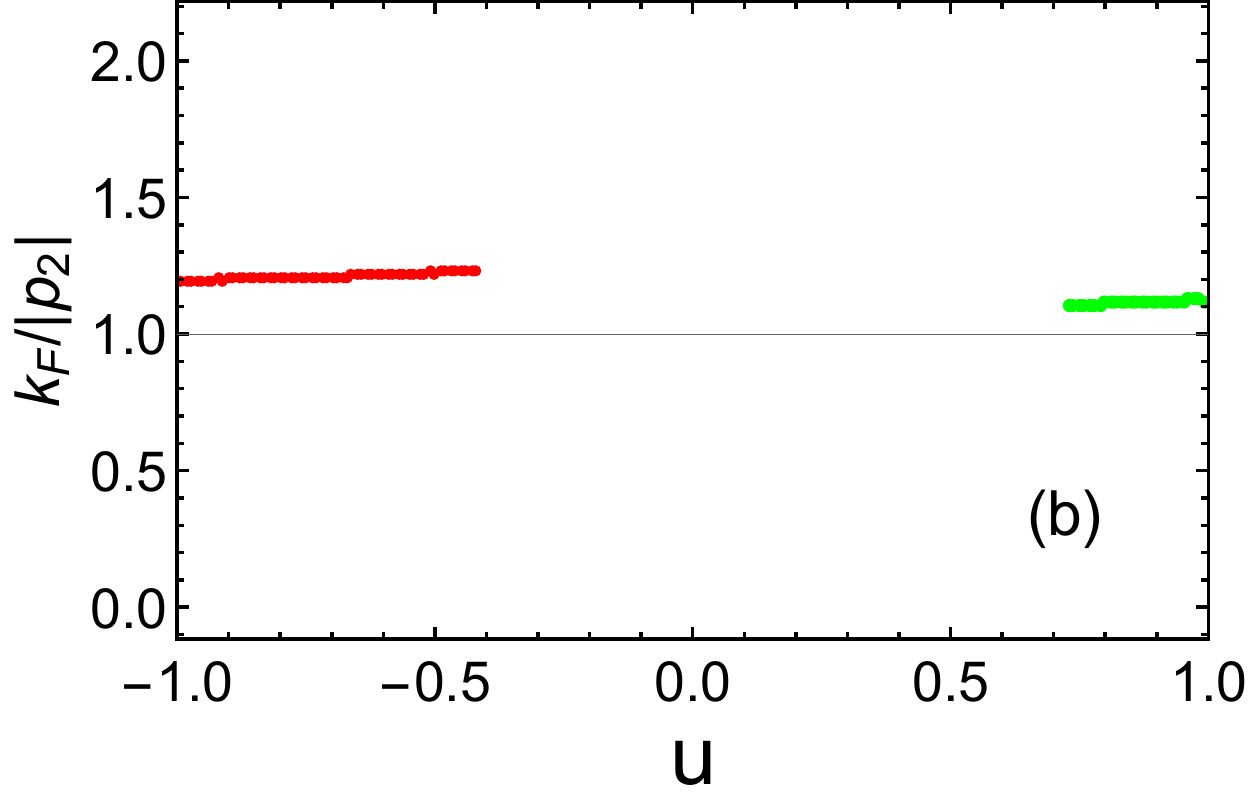}
\end{tabular}
\caption{{\small  The ratio $k_F/|p_2|$, Eq.~\eqref{210130.9}, is plotted for the minima of $\bcE$ shown in Fig.~\ref{fig.EN_minima_uv}. The panel (a) corresponds to the branches of minima with  $\bcE<0$, and the panel (b) to those with $\bcE>0$.  We observe that the ratio is always greater than 1 for all the minima pinned down.  
 \label{fig.220225.1} }} 
\end{center}
\end{figure*}

For the four branches of minima of $\bcE$ in Fig.~\ref{fig.EN_minima_uv} we check the value of the ratio $k_F/|p_2|$ in Eq.~\eqref{210130.9}, which is the relevant one for determining the  positions of the problematic poles with respect to  $k_F$. The resulting values for this ratio are plotted in Fig.~\ref{fig.220225.1}, with the panel (a) for the minima with $\bcE<0$ and the panel (b) for $\bcE>0$.
We observe from the resulting curves that $k_F/|p_2|>1$ and hence the non legitimate poles $p_2$ and $p_3$ are relevant for the densities involved in the degenerate gas of identical spin-1/2 fermions interacting in $P$-wave.
In particular, the most important one for $r_1>0(<0)$ is $p_2(p_3)$ which has positive real part.

Therefore, we conclude that a degenerate fermion gas at $T=0$ with spin-independent $P$-wave interactions, characterized by scattering volume $a_1$ and effective momentum $r_1$, is not stable.
  No acceptable minima of $\bcE$ are found  in the whole range of values for the scattering parameters, Fig.~\ref{fig.220225.1}.
 This conclusion is complementary to that reached in Ref.~\cite{Gurarie:2007} for polarized Fermi gas with identical fermions interacting in $P$-wave, so that no minima were found for normal matter. We have now extended this result also for the spin-balanced case.

 \begin{figure}[ht]
\begin{center}
\begin{tabular}{c}
  \includegraphics[width=0.45\textwidth,angle=0]{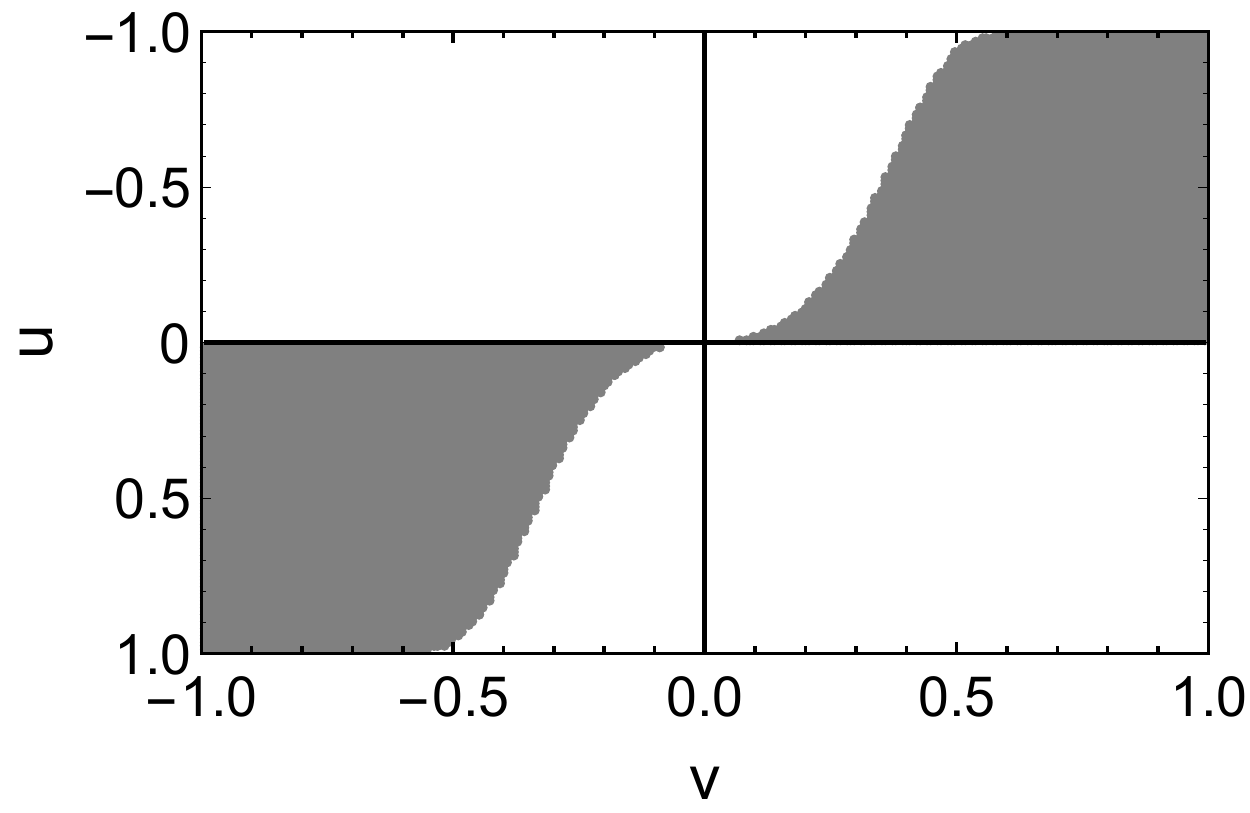}\\
\end{tabular}
\caption{{\small The shaded regions in the quadrants with $uv<0$ comprise those values of $u$ and $v$  for which $k_F/|p_2|>0.5$, such that the calculation of $\bcE$ requires to use the ERE beyond its \jo{applicability when expressed in terms of $a_1$ and $r_1$}.  The white % other
    regions %shown of the $uv$-plane
    either do not have unacceptable poles ($uv>0$) or if they do ($uv<0$) then the ratio $k_F/|p_i|<0.5$, with $|p_i|$ the absolute value of the pole position(s) of the problematic pole(s).   
 \label{fig.220317.1} }} 
\end{center}
 \end{figure}
 
 We also move on with these considerations and apply them to  the quadrants with $uv<0$ and evaluate for every value of $u$ and $v$ the ratio between $k_F$ and the modulus of the unacceptable pole(s), a ratio that we call $w(u,v)$.\footnote{It is clear from Table~\ref{tab.220130.1} that if there are more than one unallowed pole the moduli of their pole positions are common to all of them.}
 We divide the $uv$-plane in two regions as shown in Fig.~\ref{fig.220317.1}: i) The one  with $uv>0$, which are free of unacceptable poles, and those with $uv<0$ such that $w(u,v)<0.5$. The reason for this upper bound is because in the calculation of $t_m$, Eq.~\eqref{210706.1}, in-medium momenta up to $2k_F$ are involved in the function $L_m(p,a)$ of Eq.~\eqref{190624.1}. 
ii) The rest of the quadrants with $uv<0$ are shaded and for them the calculations  are based on the use of ERE out of its radius of convergence.
 Without further information  a conservative attitude requires its exclusion.
 Of course, the set of minima discussed above 
 in more detail  lie within the shaded areas of the $uv$-plane in Fig.~\ref{fig.220317.1}, because $k_F/|p_2|>1$ as already shown in Fig.~\ref{fig.220225.1}.

 Finally, we also consider the inspection of $\bcE$ as a function of $-1/a_1 k_F^3$ for $r_1=0$. The ERE $P$-wave scattering amplitude simplifies then to $k^2(-1/a_1-ik^3)^{-1}$.
 The issue here is again the raise of unacceptable poles for $a_1\geq 0$. Taking the limit $r_1\to 0$ in the expression for $p_i$, $i=1,2,3$, we have 
 \begin{align}
   p_1&=-i\frac{1}{\alpha_1}~,\\
   p_2&=\frac{\sqrt{3}+i}{2\alpha_1}~,\nn\\
   p_3&=\frac{-\sqrt{3}+i}{2\alpha_1}~.\nn
 \end{align}
 For $\alpha_1>0$ it follows that $p_1$ is a virtual state but $p_2$ and $p_3$ constitute a couple of resonance poles with positive imaginary part, which is not allowed. This settles that the radius of convergence of a well-behaved ERE should be smaller than $1/\alpha_1$ for $\alpha_1>0$.
 Therefore, in the perturbative regime in which $|a_1|\to 0$ this upper bound to the radius of convergence tends to infinity and poses no problem.
 However, as $a_1\to +\infty$ the radius of convergence tends to zero at least as fast as $1/\alpha_1$.

 Imposing then that $2k_F<1/\alpha_1$ for $a_1>0$,  the condition $8\lesssim 1/a_1k_F^3$ is required for  an ERE study  with only the scattering volume included to make sense.  
 This translates at the level of the density of the system into the upper bound $\rho<1/24 \pi^2 a_1$.

 %%%%%%%%%%%%%%%%%%%%%%%%%%%%%%%%%%%%%%%%%%%%%%%%%%%%%%%%%%%%%%%%%%%%%%%%%%%%%%%%%%%%%
 \section{Conclusions}
 \label{sec.220225.1}

 In summary, we have performed a non-perturbative renormalized calculation for unpolarized fermionic quantum liquids interacting with pure $P$-wave spin-independent interactions, which  are characterized by the scattering volume $a_1$ and the effective range $r_1$. We have then applied the novel formula of Ref.~\cite{Alarcon:2021kpx} that allows one to calculate the energy density of a spin-1/2 fermion many-body system directly in terms of the partial-wave amplitudes in vacuum by resumming the Hartree-Fock series in the ladder approximation. The minima of the energy per particle $\bcE$ are searched in the whole space of $P$-wave scattering parameters and a  rich structure of  minima is observed. Out of the four branches
 of minima  that result,  one of them  reaches values  close to (but not strictly at) the unitary limit.
 
 Further constraints are imposed on the existence of these minima by establishing whether the pole structure of the vacuum scattering amplitudes is permissible.
 It is then observed that all the observed minima in $\bcE$ correspond to scattering parameters that give rise to resonance poles with positive imaginary part in the complex momentum-plane, which is not acceptable by general arguments based on the Hermitian character of the Hamiltonian.
 Nonetheless, this is not sufficient reason to reject those minima because the ERE can give rise to artifacts beyond its radius of convergence, like generating unallowed poles, as we have exemplified with the ${^1P}_1$ and ${^3P}_1$ nucleon-nucleon $P$-wave scattering amplitudes. We have further inquired whether these problematic poles are directly affecting the interactions for the many-body system and found that this is actually the case, because  the referred poles are located with a momentum smaller than $k_F$ in absolute value.
 Then, we conclude that no stable minima result for  spin-1/2 balanced Fermi gas in normal matter. Other regions of possible values of $a_1$, $r_1$ and $k_F$ have been also identified as lying beyond the radius of expansion of the ERE.

 In addition, we have studied the energy per particle of this system around the unitary limit and given values for the leading universal parameters in an expansion in powers of $k_F$.  
 It would be also interesting to apply these methods to other systems such as unbalanced-spin quantum liquids and/or interacting with higher partial waves.  
 
\section*{Acknowledgements}

We are grateful to Manuel Valiente for bringing this problem to our attention, for many useful discussions, his participation in the writing of the introduction and a critical reading of parts of the manuscript. 
 This work has been supported in part by the  MICINN AEI (Spain) Grants PID2019-106080GB-C21/AEI/10.13039/501100011033 and  PID2019-106080GB-C22/AEI/10.13039/501100011033. 

%%%%%%%%%%%%%%%%%%%%%%%%%%%%%%%%%%%%%%%%%%%%%%%%%%%%%%%%%%%%%%%%%%%%%%%%%%%
\appendix

%%%%%%%%%%%%%%%%%%%%%%%%%%%%%%%%%%%%%%%%%%%%%%%%%%%%%%%%%%%%%%%%%%%%%%%%%%%%%%%
\section{Properties of the minima}
\label{app.220317.1}
\setcounter{equation}{0}
\def\theequation{\Alph{section}.\arabic{equation}}

\begin{figure*}[ht]
\begin{center}
  \hspace{-0.4cm}\includegraphics[width=.47\textwidth]{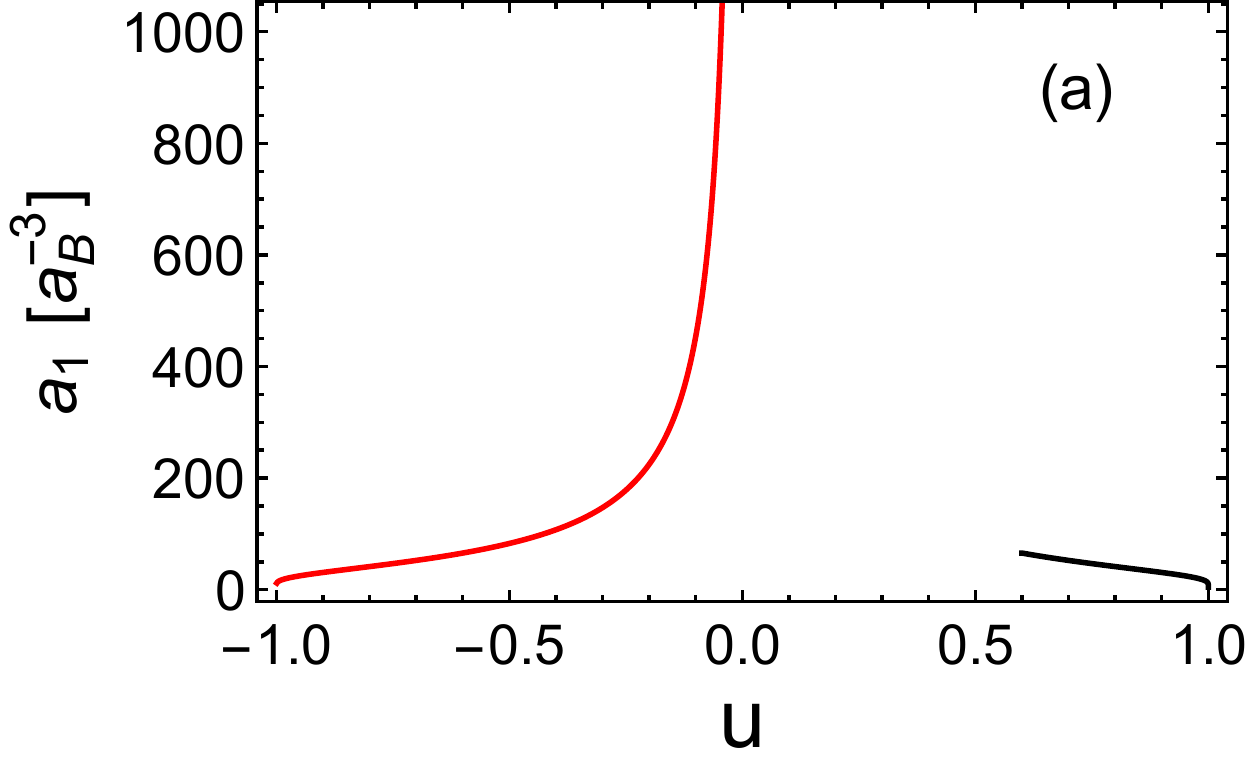}  \includegraphics[width=.45\textwidth]{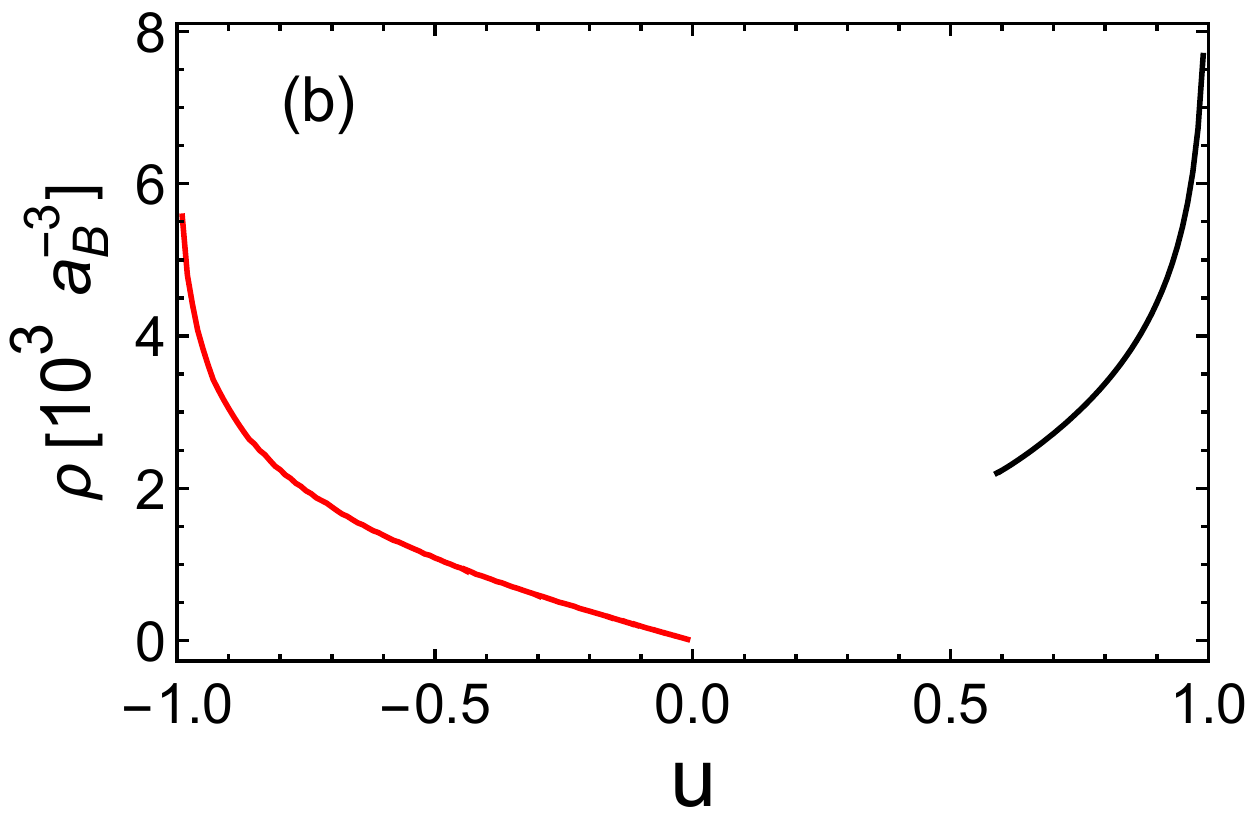}\\
    \includegraphics[width=.45\textwidth]{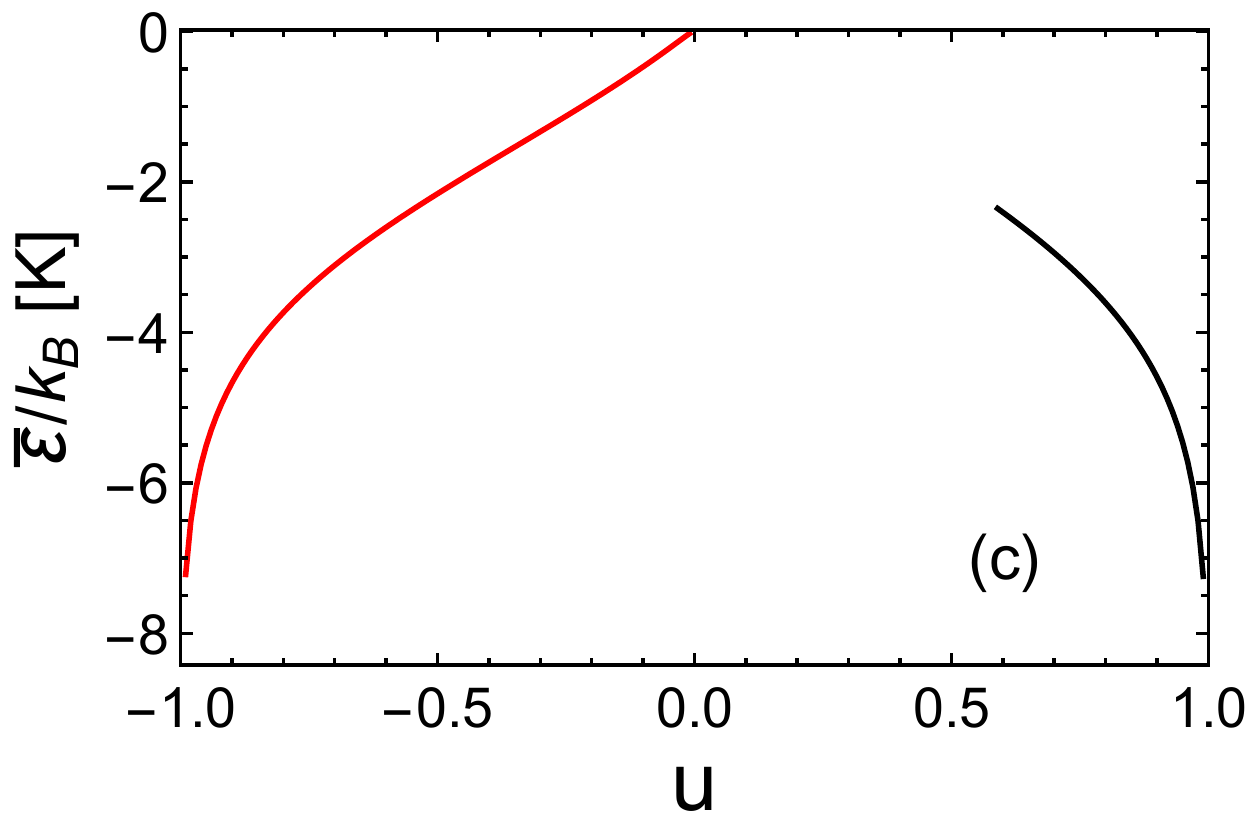}  \includegraphics[width=.45\textwidth]{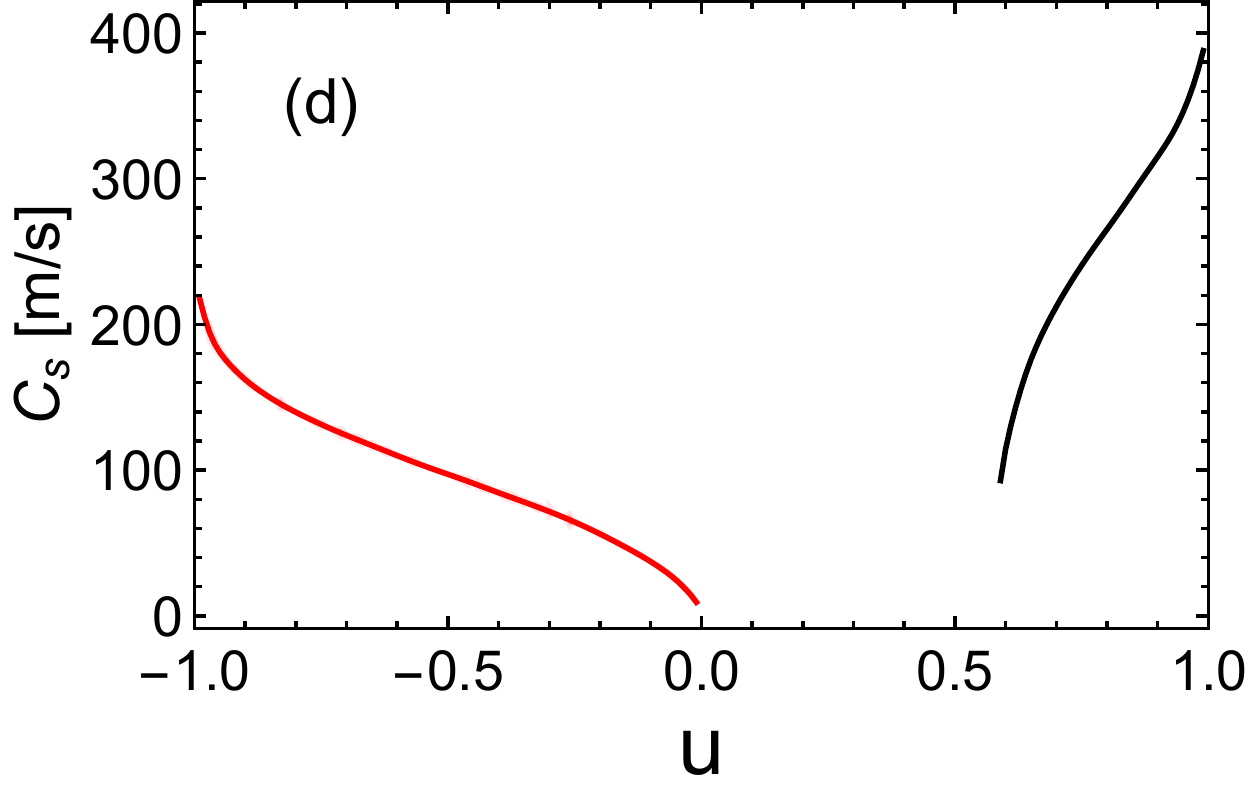}\\
      \includegraphics[width=.45\textwidth]{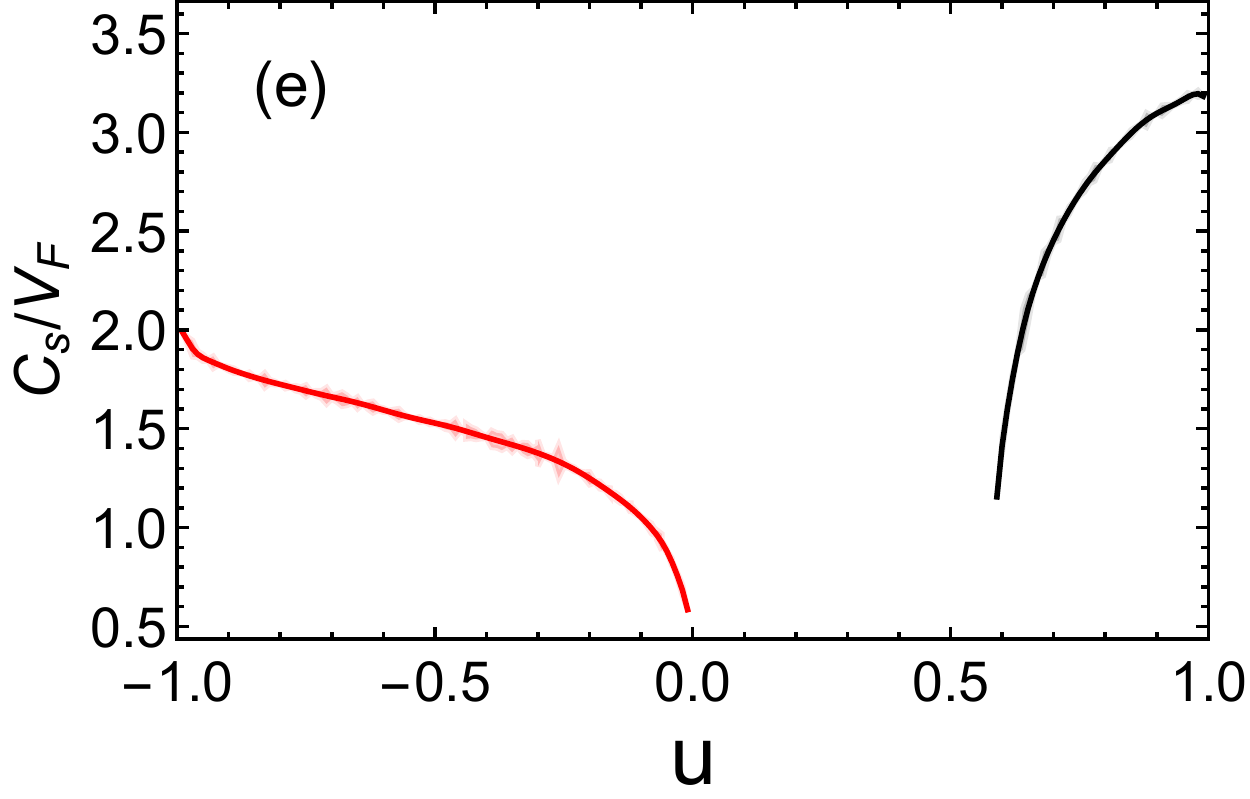}  \includegraphics[width=.45\textwidth]{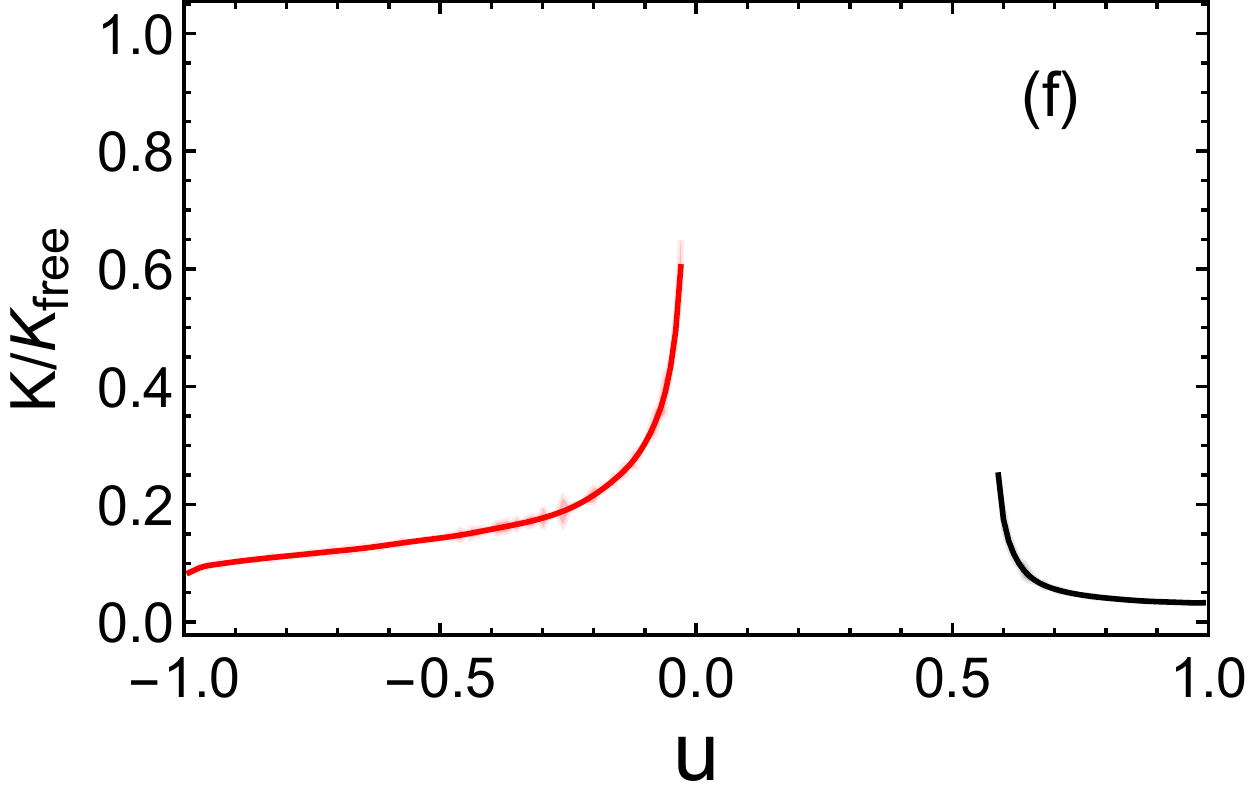}
  \caption{{\small Properties for the minima with $\bcE<0$. 
      From top to bottom and left to right  $a_1\,[a_B^{-3}]$, $\rho\,[a_B^{-3}]$, $\bcE\/k_B [K]$, $c_s\,[m/s]$,
      $c_s/v_F$ and $K/K_{\text{free}}$ are plotted.}
 \label{fig.prop.enlt0} }
\end{center}
\end{figure*}

\begin{figure*}[ht]
\begin{center}
  \hspace{-0.4cm}\includegraphics[width=.47\textwidth]{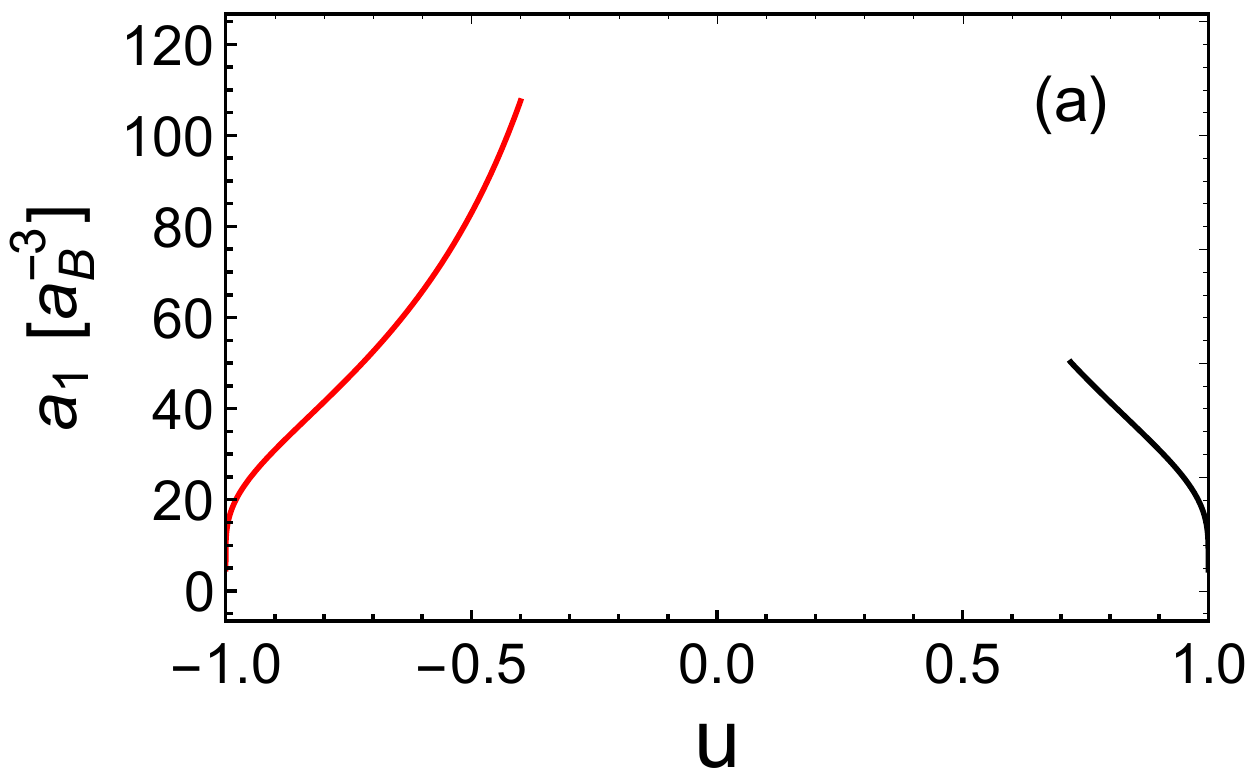}  \includegraphics[width=.47\textwidth]{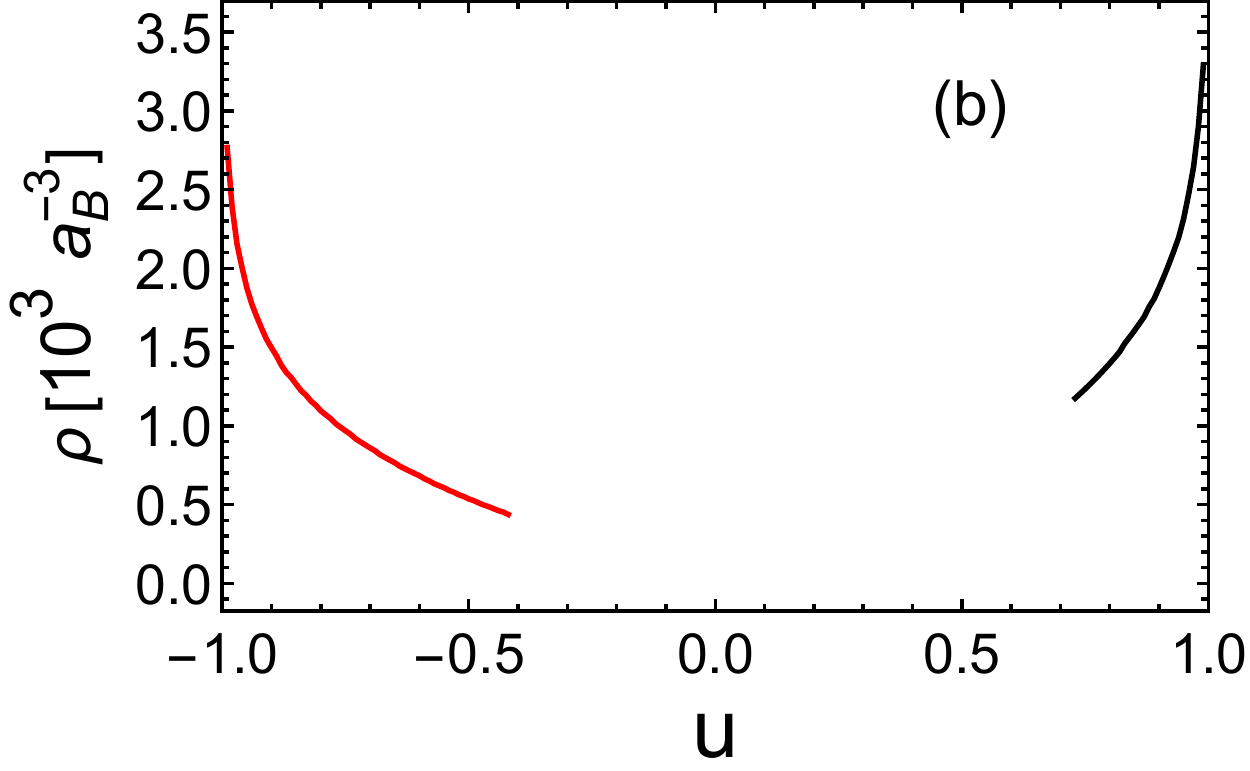}\\
    \includegraphics[width=.45\textwidth]{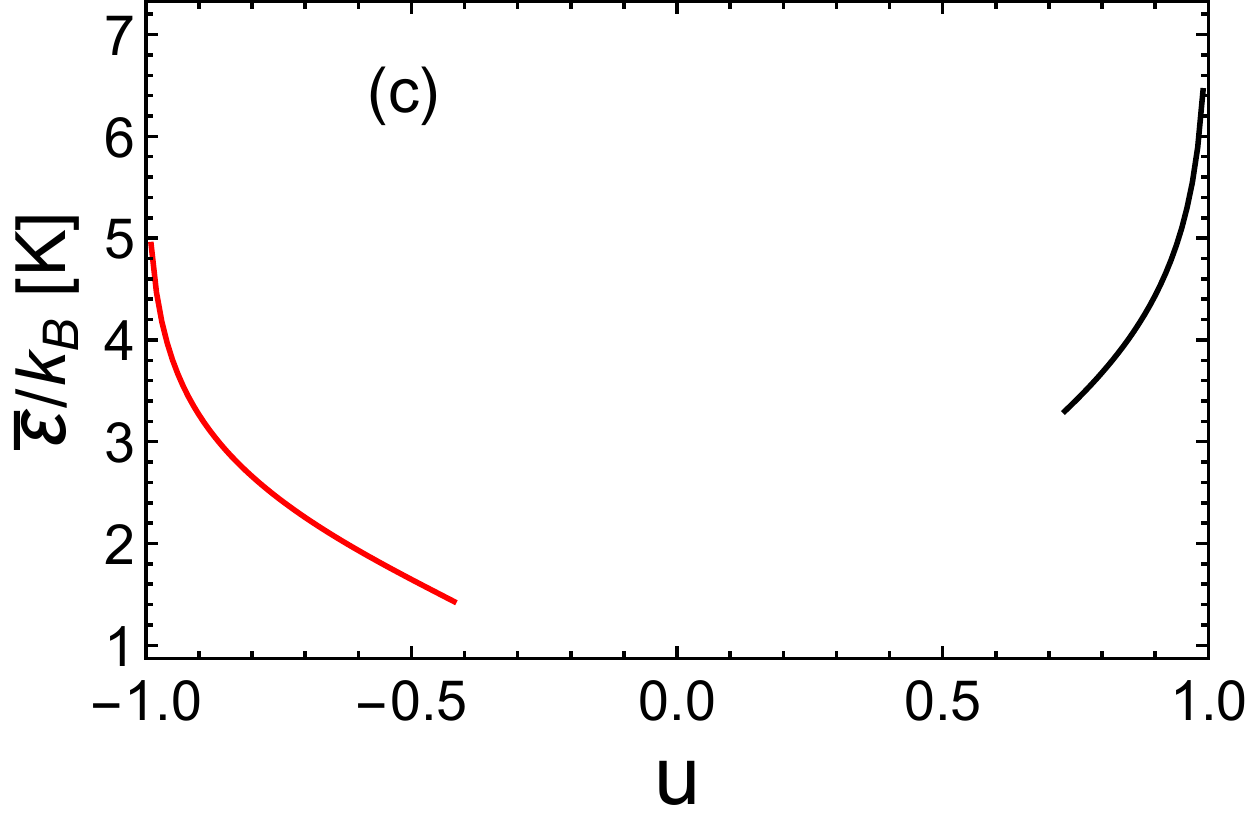}  \includegraphics[width=.47\textwidth]{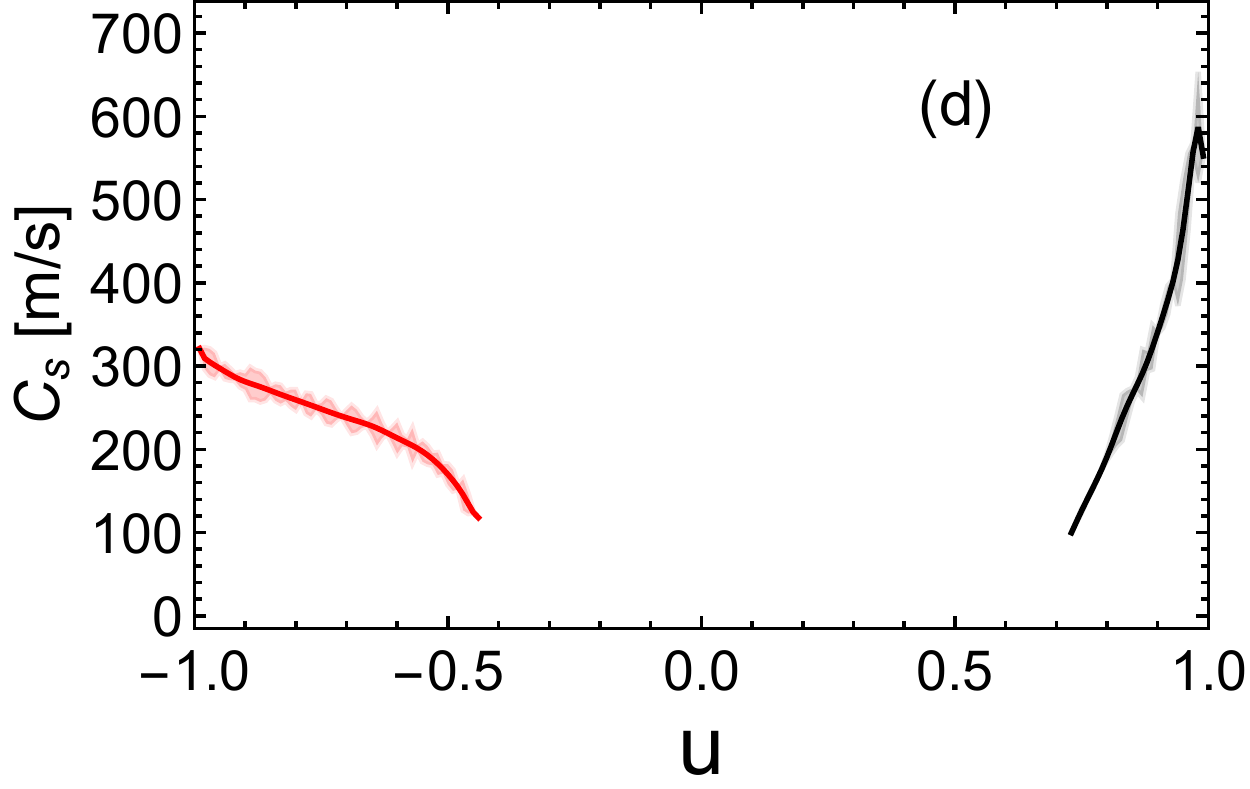}\\
      \includegraphics[width=.45\textwidth]{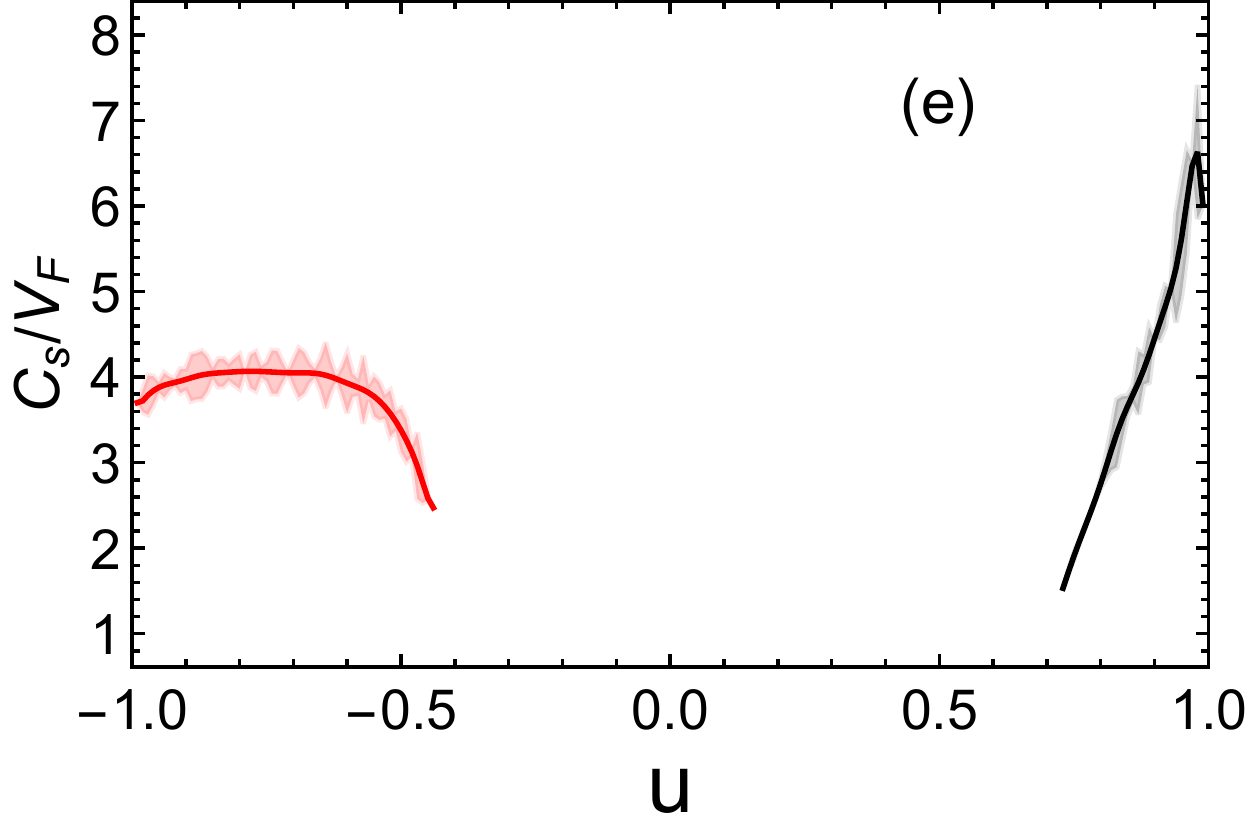}  \includegraphics[width=.48\textwidth]{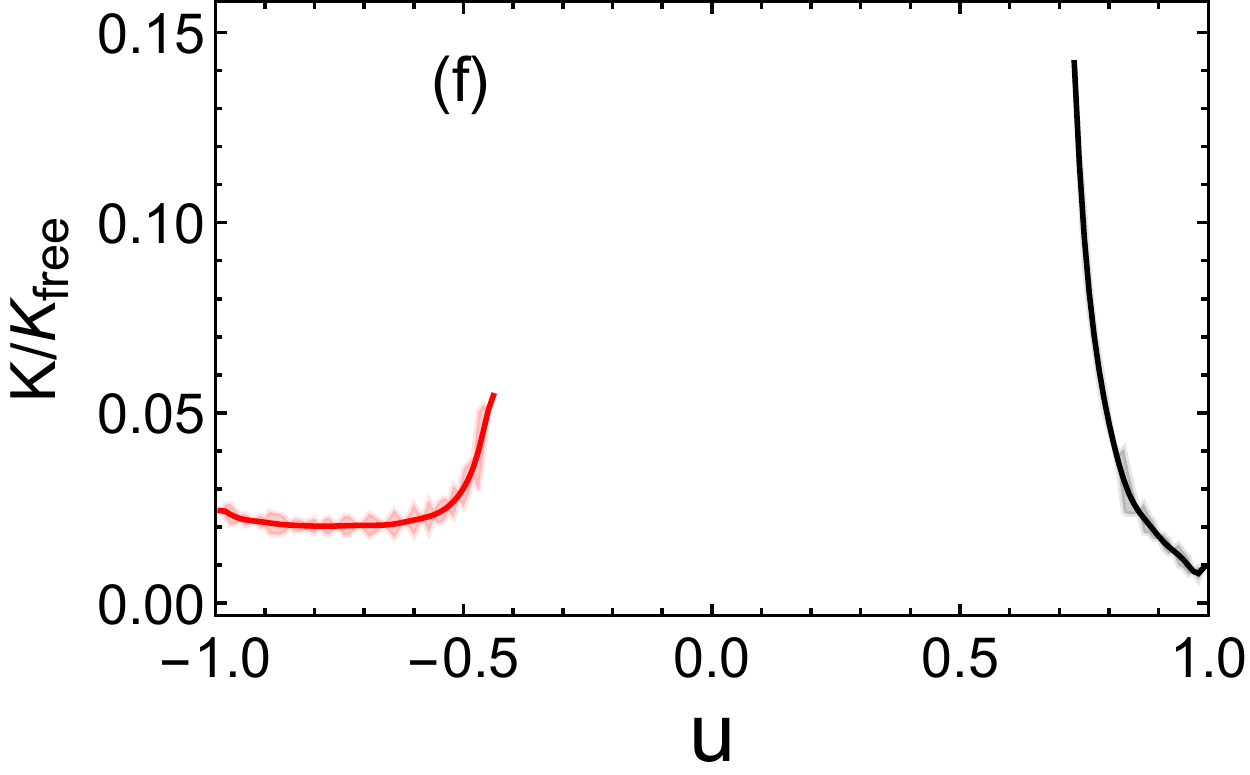}
  \caption{{\small Properties for the minima with $\bcE>0$. The light shaded areas reflect the remaining numerical uncertainty in our calculation. For the rest see the caption in Fig.~\ref{fig.prop.enlt0}. 
    }  \label{fig.prop.engt0} }
\end{center}
\end{figure*}

From the calculation of $\bcE$ we can obtain other thermodynamic and mechanical properties of the minima by employing well-known
thermodynamic identities. 
 At the minimum the pressure  $P$ 
 vanishes, while the  chemical potential $\mu$ is simply $\bcE$.  
 We also consider the sound velocity $c_s$ and the coefficient of compressibility $K$. At a minimum of $\bcE$ these magnitudes fulfill the relations
\begin{align}
\label{210605.1}
c_s^2&=\frac{\partial P}{\partial \rho(m+{\bcE})}
=
\frac{1}{m+{\bar{\cE}}}\frac{\partial}{\partial \rho}\left(\rho^2\frac{\partial {\bcE}}{\partial \rho}\right)\, ,\\
K^{-1}&=c_s^2\rho(m+\bcE)\, .\nn
\end{align}
Another interesting relation is that $K^{-1}=\rho^2\frac{\partial\mu}{\partial \rho}$, so that the derivative of the chemical potential with respect to the density is positive at the minima.  
The calculated properties of the minima are plotted in Figs.~\ref{fig.prop.enlt0} ($\bcE<0$) and
\ref{fig.prop.engt0} ($\bcE>0$), calculated with
the sign of $r_1$ chosen such that $r_1 v>0$, because then $k_F=r_1\arctanh v>0$ as it must (notice that for the minima $uv<0$).    
From top to bottom and left to right, we plot $a_1$, $\rho$ (both in units of $a_B^{-3}$), $\bcE$ in kelvin, 
$c_s$ in $m/s$, the ratio $c_s/v_F$, where $v_F=k_F/m$ is the Fermi velocity, and $K/K_{\rm free}$, with
$K_{\rm free}=\frac{k_F^5}{9\pi^2}$ the compressibility coefficient for a free Fermi gas.

In Fig.~\ref{fig.prop.enlt0} the minima with $\bcE<0$ are considered.
The left branch  when approaching  the unitary limit  has a clear gas behavior with diverging $K$ and  vanishing $\rho$, $\bcE$ and $c_s$ in that limit.  
The other branch has a maximum value for $a_1$ of $67.2\,a_B^{-1}$.
For the  branches with $\bcE>0$ drawn in Fig.~\ref{fig.prop.engt0}  their scattering volumes are also bounded from above with the maxima values of $a_1=101.8\,a_B^{-1}$ for $u<0$, and $49.0\,a_B^{-1}$ for $u>0$.
In both figures  we observe that the density of particles has
typical values of a few fermions in a volume of $10^3\,a_B^3$.
The sound velocity shows typical values of hundreds of $m/s$ and, except in the case of diverging $a_1$, it
is larger than $v_F$.

%%%%%%%%%%%%%%%%%%%%%%%%%%%%%%%%%%%%%%%%%%%%%%%%%%%%%%%%%%%%%%%%%%%%%%%%%%%%%%%
\section{Pole content as a function of $t$ and $r_1$}
\label{app.220618.1}
\setcounter{equation}{0}
\def\theequation{\Alph{section}.\arabic{equation}}

We now analyze in detail the pole content from Eq.~\eqref{220224.1} as a function of $t$ and $r_1$, indicating the regions that are allowed and excluded according whether the properties of these poles are allowed or not.

%------------------------------------------
\subsection{Pole content for $t=-1/2^{\frac{1}{3}}$}

At $t=-1/2^{\frac{1}{3}}$ it follows from Eq.~\eqref{220129.6} that 
\begin{align}
  \label{210129.11}
  r_1\alpha_1%=-3\cdot 2^{1/3}
  =-54^{1/3}~.
\end{align}
The resulting pole positions are
\begin{align}
  \label{210129.9}
  p_1&=-\frac{ir_1}{3}~,\\
  p_2&=p_1=-\frac{ir_1}{3}~,\nn\\
  p_3&=\frac{ir_1}{6}~,\nn\\
\end{align}

  {\bf Region excluded} $r_1<0$: we have a double bound-state pole which is {\bf excluded} based on general considerations \cite{Hu:1948zz,Demkov:1966}.

 {\bf Region allowed} $r_1>0$: we have a double virtual-state pole and a bound state pole ($p_3$). The residue of the bound state is

\begin{align}
  \label{210129.10}
\tau_3&=\frac{8}{r_1^2}>0~,
\end{align}

 The borderline value  $r_1=0$ is excluded because all poles are degenerate $p_i=0$ and they could be considered as a  higher-order bound-state pole. 

%------------------------------------------
\subsection{Pole content for $t<-1/2^{\frac{1}{3}}$}
\label{sec.220130.1}

%As we derived in Ref.~\cite{Alarcon:2021kpx}

In this case $|z(t)|=1$ because
\begin{align}
  \label{210929.12}
z(t)&=\frac{t}{(1+t^3+i\sqrt{-1-2t^3})^{1/3}}~,\\
|z(t)^3|&=\sqrt{\frac{t^6}{(1+t^3)^2-1-2t^3}}=1~, \nn
\end{align}
and  we  write $z(t)$ in terms of its phase $\phi(t)$, 
\begin{align}
\label{220225.1}
  z(t)&\equiv e^{i\phi(t)}~,~t<-1/2^{1/3}~.
\end{align}
For $t\to -\infty$ then $u\to e^{i2\pi/3}$ and for $t\to -1/2^{1/3}$ \jo{one has} $u\to -1$. 
\jo{Then, its} phase moves from $\phi=\pi$ ($t\to -1/2^{1/3}$)  to $\phi=2\pi/3$ ($t\to-\infty$). % (this can be seen explicitly in a numerical way too).

Let us now analyze the different pole positions. For the pole position $p_1$ we have 
\begin{align}
  \label{210929.13}
  p_1=-\frac{ir_1}{6}(1+2\cos\phi)~.
\end{align}
For $\phi=2\pi/3$ we have $1+2\cos\phi=0$. For larger values of $\phi<\pi$ then
\begin{align}
  \label{210929.14}
  1+2\cos\phi<1+2\cos2\pi/3=0~.
  \end{align}
Thus, if $r_1> 0$ then $p_1$ is a bound state, otherwise ($r_1<0$) it is a virtual state.
Regarding $\tau_1$ we have
\begin{align}
  \label{210929.15}
  \tau_1&=\frac{24}{r_1^2(1+2\cos2\phi)}>0~.
  \end{align}
This is so because $4\pi/3<2\phi<2\pi$ and then $0< 1+2\cos2\phi$.

Pole position $p_2$: 
\begin{align}
  \label{210929.16}
  p_2=\frac{ir_1}{6}(-1+\cos\phi-\sqrt{3}\sin\phi)~.
\end{align}
For $\pi<\phi<2\pi/3$ it is trivial to conclude that  $-1+\cos\phi-\sqrt{3}\sin\phi<0$ and then $p_2$ is a bound state for $r_1<0$ and a virtual state for $r_1>0$. Regarding its residue,
\begin{align}
  \label{210929.17}
  \tau_2&=\frac{12 {\rm cosec}^2\phi}{r_1^2(1-\sqrt{3}\cot\phi)}>0~,
\end{align}
because $\cot\phi<0$ for $\pi<\phi<2\pi/3$.

Pole position $p_3$:
\begin{align}
  \label{210130.1}
  p_3&=\frac{ir_1}{6}(-1+\cos\phi+\sqrt{3}\sin\phi)~.
\end{align}
For $\phi=\pi$ then $-1+\cos\phi+\sqrt{3}\sin\phi=-2$. As $\phi$ decreases towards the limit value of $2\pi/3$ then $\cos\phi+\sqrt{3}\sin\phi$ becomes larger since both $\cos\phi$ and $\sin\phi$ increases and for $\phi=2\pi/3$ one has that $-1+\cos\phi+\sqrt{3}\sin\phi=0$. Therefore, $-1+\cos\phi+\sqrt{3}\sin\phi<0$ for $\pi<\phi<2\pi/3$. Thus, $p_3$ is bound state for $r_1<0$ and a virtual state for $r_1>0$. The $p_3$ pole is shallower than $p_2$.  Regarding $\tau_3$,
\begin{align}
  \label{210130.2}
  \tau_3&=\frac{12 {\rm cosec}^2\phi}{r_1^2(1+\sqrt{3}\cot\phi)}<0~,
\end{align}
since       $1+\sqrt{3}\cot\phi<0$ for $2\pi/3<\phi<\pi$, being zero for $\phi=2\pi/3$.

 {\bf Region Excluded:} $r_1< 0$ because $p_3$ is a bound state with negative residue. 

{\bf Region Allowed:} $r_1> 0$ since $p_3$ becomes a virtual state pole.

The borderline value $r_1=0$ should be excluded because all poles are degenerate at $p_i=0$, and they could be considered bound states.

%------------------------------------------
\subsection{Pole content for $t>-1/2^{\frac{1}{3}}$}
\label{sec.220130.1}

For these values of $t$ then $z(t)\in \mathbb{R}$. 
We have that $z<0$ for $-1/2^{1/3}<t<0$ and $z>0$ for $t>0$.

Pole position $p_1$: 
\begin{align}
  \label{210130.3}
  p_1&=-\frac{i r_1}{6}(1+z+z^{-1})~.
\end{align}
For $-1/2^{1/3}<t<0$, i.e. $-1<z<0$, then $h(z)\equiv 1+z+z^{-1}$ has values within the interval $(-\infty,-1)$, while for $0<t$, i.e. $0<z$, then $h(z)$ decreases from $+\infty$ to 0 as $z$ increases.
 As a result, $p_1$ is either a bound state or a virtual state depending on the range of values of $t$ for a given $r_1$ according to:

a) $-1/2^{1/3}<t<0$:  $p_1$ is a bound state for $r_1>0$, and a virtual state if $r_1<0$. 

b) $0<t$:  $p_1$ is a virtual state for $r_1>0$, while  $p_1$ is a bound state when $r_1<0$.

Regarding its residue $\tau_1$\, ,
\begin{align}
  \label{210130.4}
  \tau_1&=\frac{24 z^2}{r_1^2(1+z^2+z^4)}>0 
\end{align}
for non-vanishing $z$, and no extra constraint stems from here.

The poles $p_2$ and $p_3$ are resonance poles that share the same imaginary part and have real parts with different sign. Their imaginary and real parts are, respectively, 
\begin{align}
  \label{210130.5}
        {\rm Im}p_2&={\rm Im}p_3=\frac{r_1}{12 z}(1-z)^2~,\\
        {\rm Re}p_2&=-{\rm Re}p_3=\frac{r_1\sqrt{3}(1-z^2)}{12 z}~.\nn
\end{align}
 Then, in the range $-1/2^{1/3}<t<0$ ($z>0$) we have  ${\rm Im}p_2<0$ for $r_1>0$, while ${\rm Im}p_2>0$ if $r_1<0$, which is {\bf excluded}. 
 In turn, for $0<t$ ($0<z$) it follows that ${\rm Im}p_2>0$ for $r_1>0$, which is {\bf excluded}, while ${\rm Im}p_2<0$ for $r_1<0$.

%%%%%%%%%%%%%%%%%%%%%%%%%%%%%%%%%%%%%%%%%%%%%%%%%%%%%%%%%%%%%%%%%%%%%%%%%%%%%%%%%%%%%%%%%
\bibliography{unitary.bib}
\bibliographystyle{apsrev4-1}

\end{document}